\documentstyle[preprint,prd,aps,eqsecnum,epsf]{revtex}

\begin{document}

\preprint{VAND-TH-98-02
\hspace{-34.5mm}\raisebox{-2.4ex}{DART-HEP-98/02}
\hspace{-3.1cm}\raisebox{-4.8ex}{IF/UERJ-98/10}
\hspace{-2.7cm}\raisebox{-7.2ex}{March 1998}
\hspace{-2.7cm}\raisebox{-9.6ex}{hep-ph/9803394}}

\title{Strong Dissipative Behavior in Quantum Field Theory}

\author{Arjun Berera$^{1}\;$\thanks{E-mail:
berera@vuhep.phy.vanderbilt.edu}$\;$,
Marcelo Gleiser$^{2}\;$\thanks{E-mail: gleiser@peterpan.dartmouth.edu.
$\;$ NSF Presidential Faculty Fellow.}$\;$ and
Rudnei O. Ramos$^{3}\;$\thanks{E-mail: rudnei@symbcomp.uerj.br}}

\address{
{\it $^1\;$ Department of Physics and Astronomy, Vanderbilt University,}\\
{\it Nashville, TN 37235, USA}
\\
{\it $^2 \;$ Department of Physics and Astronomy, Dartmouth College,}\\
{\it Hanover, NH 03755, USA}
\and \\
{\it Nasa/Fermilab Astrophysics Center, Fermi National Accelerator
Laboratory}\\
{\it Batavia, IL 60510, USA}
\and \\
{\it Osservatorio Astronomico di Roma}\\
{\it Vialle del Parco Mellini 84, Roma I-00136, Italy}
\\
{\it $^3 \;$ Universidade do Estado do Rio de Janeiro,}
{\it Inst. de F\'{\i}sica - Depto. de F\'{\i}sica Te\'orica,}\\
{\it 20550-013 Rio de Janeiro, RJ, Brazil}}

\maketitle

\bigskip
In Press Physical Review D, 1998

\bigskip
\bigskip
\begin{abstract}

We study the conditions under which an overdamped regime can be attained in
the dynamic evolution of a quantum field configuration. Using a
real-time formulation of finite temperature field theory, we compute the
effective evolution equation of a scalar field configuration,
quadratically interacting with a given set of other scalar fields. We
then show that, in the overdamped regime, the dissipative kernel in the
field equation of motion is closely related to the shear viscosity
coefficient, as computed in scalar field theory at finite temperature.
The effective dynamics is equivalent to a time-dependent Ginzburg-Landau
description of the approach to equilibrium in phenomenological theories
of phase transitions.
Applications of our results, including a recently proposed inflationary
scenario called ``warm inflation'', are discussed.

\vspace{0.34cm}
\noindent
PACS number(s): 98.80 Cq, 05.70.Ln, 11.10.Wx

\end{abstract}

\section{Introduction}

Kinetic equations describe the time evolution of a certain chosen set of
physical variables. The choice of physical variables in principle is
arbitrary, but often in practice is governed by the measurement of
interest. Typical examples are the order parameter of a complex system
or the coordinate of a Brownian particle in a heat reservoir. The
kinetic approach is usually implemented through a proper separation of
the microscopic equations of motion of the chosen physical variables
into regular and random parts. An averaging over the random part then
generates the effective partition function for the regular part. This
averaging is often referred to as a coarse-graining.

One typical application of the kinetic approach is when the physical
variables of interest possess energy in relative excess or deficiency to
the rest of a large system. Kinetic theory then describes the approach
to equilibrium of the chosen physical variables, as for example in the
kinetics of phase transitions or in Brownian motion. In the former case,
the system is able to release energy to the environment due to some
change in its internal state. Provided the environment is
disproportionately large relative to the system, the process is
irreversible. For a continuous transition, the focus of the present
work, this process of equilibration can be described by the monotonic
change of an appropriate order parameter, which is the chosen physical
variable. Many systems are known to relax in this manner.
Phenomenologically, they are successfully described by the
time-dependent Ginzburg-Landau theory \cite{gold}.

Here we are interested in examining under which circumstances physical
variables whose microscopic dynamics is second order in time, as for
example the Higgs order parameter of spontaneous symmetry breaking, may
have a dynamics which is effectively first order in time as in 
Ginzburg-Landau phenomenological models.

Qualitatively it is not difficult to argue the plausibility of this
standard picture for the Higgs symmetry breaking scenario. A single
variable, the Higgs order parameter, is modeled to control the release
of energy to all the modes that couple to it. By basic notions of
equipartition, one anticipates that some portion of the order
parameter's energy will flow irreversibly to any given mode. Provided
the Higgs order parameter couples to a sufficient number of modes, the
motion of the order parameter will be overdamped.

In particle physics models, Higgs symmetry breaking is accompanied by
mass generation. Thus the natural couplings for the Higgs field $\phi$
to bosonic fields $\chi_i$ is $\phi^2 \chi_i^2$, gauge fields $A^{\mu}_i$
is $\phi^2 A^{i\mu} A_{i \mu}$ and fermionic fields $\psi_i$ is $\phi
{\bar \psi}_i \psi_i$. For a microscopic realization of time dependent
Ginzburg-Landau theory for the Higgs scalar order parameter in a
particle physics setting, these are the most obvious types of couplings
to investigate. In this paper we will examine 
the case of purely bosonic couplings in the
``symmetry restored'' regime. That is, we will study the relaxation of an
order parameter which is initially away from the only minimum of the
free energy density describing the system. Much of the formalism
required for this has already been done in \cite{GR} but we will extend
that calculation to the overdamped regime. In an upcoming paper, we plan
to study the symmetry broken case.

To our knowledge, this paper is the first study of overdamping in
quantum field theory with realistic couplings between system and
environment, as inspired by particle physics. Overdamping has been
studied in quantum mechanical reaction rate theory for a particle
escaping from a metastable state (for a review please see
\cite{hangii}). This is sometimes referred to as the Kramer's problem,
with the overdamped limit also called the Smoluchowski limit. Quantum
mechanical models describing this problem are commonly of the
system-heat bath type. Microscopic quantum mechanical models have been
constructed along these lines, in which the particle (system) is coupled
to a set of otherwise free harmonic oscillators (heat bath). Such
microscopic system-heat bath models are often referred to as
Caldeira-Leggett (CL) models. In many cases they have been exactly solved
\cite{fordkac}. The overdamped limit has been derived in these models
for the case where the coupling is linear with respect to the oscillator
variables but arbitrary with respect to the particle variable
\cite{hangii,cl2}.

A Caldeira-Leggett type model has also been formulated for the case
where the system is a self interacting scalar quantum field 
coupled linearly to a set of otherwise free fields and the overdamped
limit has been obtained \cite{ab54}. This model does provide a
microscopic quantum mechanical realization of time-dependent
Ginzburg-Landau dynamics in scalar quantum field theory. However, since
the couplings between system and environment variables are linear, it
should be considered as a first step toward more realistic treatments.
More importantly, the calculational method used in \cite{ab54} cannot be
extended to the case when the system variable couples quadratically
to other fields.

Although the analysis of overdamping in this paper has general
applicability, it was motivated by the warm inflation scenario of the
early universe \cite{ab54,wi}. In \cite{wi} it was realized that the
standard Higgs symmetry breaking scenario, when put into a cosmological
setting, provides suitable conditions for the universe to enter a de
Sitter expansion phase and then smoothly exit into a radiation dominated
phase. The overdamped motion of the order parameter in this scenario may
sustain the vacuum energy sufficiently long for de Sitter expansion to
solve the horizon and flatness problems. Simultaneously, the
relaxational kinetics of the order parameter can maintain the
temperature of the universe and permit a smooth exit from the de Sitter
phase into the radiation dominated phase. Finally, the thermal
fluctuations of the order parameter provide the initial seeds of density
perturbations, which in addition could be scale free under specified
conditions \cite{wi,bf2}. An elementary analysis of this scenario, based
on Friedmann cosmology for general realizations of order parameter
kinematics, indicated that if the universe's
temperature does not fall too much during de Sitter expansion,
then the cosmological
expansion factor from the de Sitter phase should be of order the lower bound
set by observation \cite{ab55}. Although this is not a tight constraint
of this scenario, it is a natural one. An analysis of COBE data
motivated by this expectation did indicate a slight preference for a
small super-Hubble suppression scale, which could be interpreted as
arising from a de Sitter expansion with duration near its lower bound
\cite{bfh}. Furthermore, the overdamped limit required by warm
inflation, when expressed in different terms, was noted \cite{ab54} to
be an adiabatic limit, for which known methods from dissipative quantum
field theory \cite{GR,hs1,morikawa} are presumed valid. These facts provide
further motivation to seek a microscopic model of the scenario, which is
the goal of the present work.

The calculational methods used here, based on Schwinger's
close-time path formalism, were developed in \cite{GR}. There are
several other works in the literature that apply this formalism to a
variety of different situations. [See, for example, the works of Refs.
\cite{hu,boya1,morikawa,greiner,yoko}.] The new feature of the present
paper is to shift focus to a kinematic regime dominated by strong
dissipation, in order to establish under which conditions this regime
leads to overdamped motion. This approach will allow us to have a unique
understanding of the microphysical origin of such dynamical behavior,
which is in general invoked phenomenologically in applications ranging
from condensed matter physics to inflationary cosmology.

The paper is organized as follows. In Sec. II our model of interacting
bosons is presented and the effective action is computed perturbatively
for a homogeneous time dependent background field configuration
$\phi(t)$. In Sec. III the effective Langevin-like equation of motion
is obtained for $\phi$ in the symmetry-restored phase. In Sec. IV the
overdamped limit of this equation of motion is derived and regions of
validity are given. In Sec. V the results of the previous sections,
which are for Minkowski space, are extrapolated into a cosmological
setting and a preliminary examination is made of the warm inflation
scenario. In Sec. VI concluding remarks are given. Two Appendixes 
are included
to clarify a few
technical details, like the evaluation of the imaginary part of
the self-energies and to stress
the importance of taking fully-dressed field propagators to properly
describe dissipation in the adiabatic approximation for the field 
configuration.


\section{Model of Interacting Bosonic Fields}

\subsection{The Effective Action}

Let us consider the following model of a scalar field $\phi$
in interaction with $N$ scalar fields $\chi_j$:

\begin{equation}
{\cal L}[\phi,\chi_j] = \frac{1}{2}
(\partial_\mu \phi)^2 - V[\phi] + \sum_{j=1}^N \left\{
\frac{1}{2} (\partial_\mu \chi_j)^2 - V[\chi_j] \right\} -
V_{\rm int} [\phi,\chi_j]\: ,
\label{Nfields}
\end{equation}

\noindent
where

\begin{equation}
V[\phi] = \frac{m^2}{2}\phi^2 + \frac{\lambda}{4 !} \phi^4 \;,
\label{Vphi}
\end{equation}

\begin{equation}
V[\chi_j] = \frac{\mu_j^2}{2}\chi_j^2 + \frac{f_j}{4!} \chi_j^4  \:,
\label{Vchi}
\end{equation}

\noindent
and

\begin{equation}
V_{\rm int}[\phi,\chi_j] = \sum_{j=1}^N \frac{g_j^2}{2} \phi^2
\chi_j^2 \:.
\label{Vint}
\end{equation}

\noindent
{}For the most part, we
will consider all coupling constants positive: $\lambda$,
$f_j$ and $g_j^2$ $> 0$.
Writing $\phi \to
\varphi + \eta$ in (\ref{Nfields}), where $\varphi$ is a background 
field configuration and $\eta$ are small fluctuations around $\varphi$,
we obtain the
expression for the 1-loop  effective action $\Gamma
[\varphi]$, valid to second order in the fluctuations, 
by performing the functional (Gaussian) integrations
in $\eta$  and $\chi_j$:

\begin{equation}
\Gamma[\varphi] = S[\varphi] +
\frac{1}{2} i {\rm Tr} \ln \left[ \Box + V''(\varphi)
\right] + \frac{1}{2} i \sum_{j=1}^N
{\rm Tr} \ln \left[ \Box + \mu_j^2 + g_j^2
\varphi^2 \right] \:,
\label{action1}
\end{equation}

\noindent
where $S[\varphi] = \int d^4 x {\cal L}[\varphi,0]$,
$V''(\varphi) = \frac{d^2}{d \phi^2} V[\phi] \Big|_{_{\phi =
\varphi}}$, and

\begin{eqnarray}
\lefteqn{\frac{1}{2} i {\rm Tr} \ln \left[\Box + V''(\varphi)\right] +
\frac{1}{2} i \sum_{j=1}^N
{\rm Tr} \ln \left[ \Box + \mu_j^2 + g_j^2 \varphi^2 \right]=}
\nonumber \\
& & = - i \ln \int D \eta \prod_j D \chi_j
\exp \left\{ - \frac{i}{2} \eta \left[
\Box + V''(\varphi) \right] \eta - \frac{i}{2} \chi_j \left[
\Box + \mu_j^2 + g_j^2 \varphi^2 \right] \chi_j \right\} \: .
\label{Trln}
\end{eqnarray}

Neglecting contributions to (\ref{action1}) which are independent of
$\varphi$, we can expand the
logarithms in (\ref{action1}) in powers of $\varphi$, obtaining,
in the graphic representation:

\begin{equation}
\begin{picture}(270,28)

\put(-85,0){$\Gamma[\varphi]_{_{\rm 1-loop}}=  $}

\thicklines

\put(0,0){\line(-1,-1){10}}
\put(10,0){\circle{20}}
\put(0,0){\line(-1,1){10}}
\put(7,15){$\phi$}

\put(25,0){+}

\put(50,0){\line(-1,-1){10}}
\put(60,0){\circle{20}}
\put(50,0){\line(-1,1){10}}
\put(70,0){\line(1,1){10}}
\put(70,0){\line(1,-1){10}}
\put(57,15){$\phi$}
\put(57,-20){$\phi$}

\put(85,0){+}
\put(100,0){$\sum_{j=1}^N \Big[$}

\put(150,0){\line(-1,-1){10}}
\put(150,0){\line(-1,1){10}}
\put(160,0){\circle{20}}
\put(157,15){$\chi_j$}

\put(175,0){+}

\put(200,0){\line(-1,-1){10}}
\put(210,0){\circle{20}}
\put(200,0){\line(-1,1){10}}
\put(220,0){\line(1,1){10}}
\put(220,0){\line(1,-1){10}}
\put(207,15){$\chi_j$}
\put(207,-20){$\chi_j$}

\put(240,0){$\Big]$}

\put(245,0){$+\: {\cal O}(\lambda^3) + {\cal O}(g_j^6)\: ,$}

\end{picture}
\label{graphs1l}
\end{equation}

\vspace{0.5cm}

\noindent
where we have identified the propagators in the internal lines.
External lines are $\lambda \varphi^2/2$ for the $\phi$-graphs
and $g_j^2 \varphi^2$ for the $\chi$-graphs.

\subsection{Single-Particle
Excitations and Dissipation: Dressing the Propagators}

Before presenting our derivation of the effective
nonequilibrium equation of motion for $\varphi$, we contrast
our approach with earlier works in the
literature. We closely follow the method of Ref.
\cite{GR} in the derivation of the evolution equation for
$\varphi$. In particular, it was shown in \cite{GR}
that for slowly changing fields,
dissipative terms vanish if they are computed perturbatively with bare
propagators.
There are several issues related to this result.
Boyanovsky {\it et al.} in Ref. \cite{boya1} argue, in the context of
a toy model,
that dissipative effects cannot be studied within perturbation
theory: perturbation theory breaks down
before dissipative effects can be observed. This shows that dissipation
is a nonperturbative effect in quantum field theory. In
\cite{GR} it was shown that dissipative terms can be derived once a
consistent ``dressing'' of propagators is used. This is an explicit way
of considering the effect of quasi-particles (or single-particle states)
in the evolution of the system, described by $\varphi$, in interaction with
a thermal bath which represents fluctuations of $\phi$ and of others
fields to which it may be coupled.

It seems reasonable to expect that
dissipation effects are closely related to the effect of collisions 
which dress
the field propagators. Take for example the case of a bare propagator
expressed in terms of the spectral density $\rho_0 (p)$, where there is a one
to one correspondence between the energy and the momentum of a given state.
This completely neglects the 
spreading of possible energy states due to interactions. In a full
``dressing'' of propagators, this is accounted for through the
introduction of a lifetime (decay width) for single-particle states,
such that the full (dressed) spectral density $\rho (p)$ is smeared
out. In particular, particle lifetimes are crucial in
the study of relaxation time-scales
in quantum many-body theory \cite{kada,mahan}.

Also, the reason why we can get dissipation within our approach
can be traced back to the very way that transport
coefficients are derived in quantum field theory. As we will
show later, the assumption of a slowly moving field is
consistent with overdamping
in a strong dissipative environment, justifying the adiabatic approximation
we adopted. In this regime, there is a close
relation between the dissipation we compute and the shear viscosity
computed from the Kubo formula
\cite{takao,jeon1,jeon2,heinz}. As explained in
\cite{jeon1,jeon2}, diagrams contributing to the shear viscosity have
near on-shell singularities for free bare propagators. Full resummed
propagators regulate these singularities through an explicit thermal
lifetime of single particle excitations. Analogous singularities are
exhibited by our expressions for dissipation terms if bare propagators
are used. Additional issues concerning the relation of our dissipation
terms with the shear viscosity will be discussed in the following two
sections.

\subsection{Self-Energies and Dressed Propagators}

{}From the above discussion, we rewrite the Lagrangian
density in (\ref{Nfields}) as

\begin{eqnarray}
{\cal L}  & = & \frac{1}{2} \left( \partial_\mu \phi
\right)^2 - \frac{1}{2} \left( m^2 + \Sigma_\phi \right) \phi^2 -
\frac{\lambda}{4 !} \phi^4 + \frac{1}{2} \Sigma_\phi \phi^2 +
\nonumber \\
& + & \sum_{j=1}^N \left\{
\frac{1}{2} (\partial_\mu \chi_j)^2 -
\frac{1}{2} \left(  \mu_j^2 + \Sigma_{\chi_j} \right) \chi_j^2 -
\frac{g_j^2}{2} \phi^2
\chi_j^2 - \frac{f_j}{4!} \chi_j^4 + \frac{1}{2}
\Sigma_{\chi_j} \chi_j^2 \right\}
\;,
\label{dressedL}
\end{eqnarray}

\noindent
where $\Sigma_\phi$ and $\Sigma_{\chi_j}$ are the self-energies for the
$\phi$ and $\chi_j$ fields, respectively. This way we can work with
full (dressed) propagators for the $\phi$ and $\chi_j$ fields
(note the implicit resummation of
diagrams involved in this operation) and at the same time keep
consistency by considering $\frac{\lambda}{4 !} \phi^4  -
\frac{1}{2} \Sigma_\phi \phi^2$  and also $ \frac{f_j}{4!}
\chi_j^4 - \frac{1}{2} \Sigma_{\chi_j} \chi_j^2$
as interaction terms. This method has already been adopted before in
many different contexts [see, for example \cite{GR,heinz,parw}]. In
terms of the self-energies the field propagators are written as

\begin{equation}
\frac{1}{q^2 - m^2 + i \epsilon} \longrightarrow \frac{1}{q^2 - m^2 -
\Sigma (q) + i \epsilon} \;.
\label{newprop}
\end{equation}

{}For both $\phi$ and $\chi_j$ fields, a finite lifetime of single
particle excitations, given in terms of the imaginary part of the self-
energies, first appear at the two-loop order. We thus restrict, for
simplicity, the evaluation of $\Sigma_\phi$ and $\Sigma_{\chi_j}$ up to
the two-loop level. Diagrammatically, the self-energies are given by

\vspace{0.75cm}
\[
\begin{picture}(300,28)
\put(-80,0){$\Sigma_\phi \: =$}
\thicklines
\put(-40,0){\line(1,0){40}}
\put(-20,10){\circle{20}}
\put(-23,25){$\phi$}

\put(10,0){+}

\put(30,0){\line(1,0){40}}
\put(50,10){\circle{20}}
\put(47,25){$\chi$}

\put(80,0){+}

\put(100,0){\line(1,0){40}}
\put(120,10){\circle{20}}
\put(120,30){\circle{20}}
\put(117,45){$\phi$}
\put(100,10){$\phi$}
\put(133,10){$\phi$}

\put(150,0){+}

\put(170,0){\line(1,0){40}}
\put(190,10){\circle{20}}
\put(190,30){\circle{20}}
\put(187,45){$\phi$}
\put(203,10){$\chi$}
\put(170,10){$\chi$}

\put(220,0){+}

\put(240,0){\line(1,0){40}}
\put(260,10){\circle{20}}
\put(260,30){\circle{20}}
\put(257,45){$\chi$}
\put(240,10){$\chi$}
\put(273,10){$\chi$}

\put(290,0){+}

\put(310,0){\line(1,0){40}}
\put(330,10){\circle{20}}
\put(330,30){\circle{20}}
\put(327,45){$\chi$}
\put(310,10){$\phi$}
\put(343,10){$\phi$}

\end{picture}
\]

\vspace{-0.3cm}

\begin{equation}
\begin{picture}(300,28)
\thicklines
\put(-30,0){+}

\put(-10,0){\line(1,0){60}}
\put(20,0){\circle{30}}
\put(17,20){$\phi$}
\put(17,3){{\scriptsize $\phi$}}
\put(17,-25){$\phi$}

\put(60,0){+}

\put(80,0){\line(1,0){60}}
\put(110,0){\circle{30}}
\put(107,20){$\chi$}
\put(107,3){{\scriptsize $\phi$}}
\put(107,-25){$\chi$}

\put(150,0){+}

\put(170,0){higher loop terms}
\end{picture}
\label{selfphi}
\end{equation}

\noindent
and

\vspace{1.5cm}
\[
\begin{picture}(300,28)
\put(-80,0){$\Sigma_\chi \: =$}
\thicklines
\put(-40,0){\line(1,0){40}}
\put(-20,10){\circle{20}}
\put(-23,25){$\chi$}

\put(10,0){+}

\put(30,0){\line(1,0){40}}
\put(50,10){\circle{20}}
\put(47,25){$\phi$}

\put(80,0){+}

\put(100,0){\line(1,0){40}}
\put(120,10){\circle{20}}
\put(120,30){\circle{20}}
\put(117,45){$\chi$}
\put(100,10){$\chi$}
\put(133,10){$\chi$}

\put(150,0){+}

\put(170,0){\line(1,0){40}}
\put(190,10){\circle{20}}
\put(190,30){\circle{20}}
\put(187,45){$\phi$}
\put(203,10){$\chi$}
\put(170,10){$\chi$}

\put(220,0){+}

\put(240,0){\line(1,0){40}}
\put(260,10){\circle{20}}
\put(260,30){\circle{20}}
\put(257,45){$\phi$}
\put(240,10){$\phi$}
\put(273,10){$\phi$}

\put(290,0){+}

\put(310,0){\line(1,0){40}}
\put(330,10){\circle{20}}
\put(330,30){\circle{20}}
\put(327,45){$\chi$}
\put(310,10){$\phi$}
\put(343,10){$\phi$}

\end{picture}
\]

\vspace{-0.3cm}

\begin{equation}
\begin{picture}(300,28)
\thicklines
\put(-30,0){+}

\put(-10,0){\line(1,0){60}}
\put(20,0){\circle{30}}
\put(17,20){$\chi$}
\put(17,3){{\scriptsize $\chi$}}
\put(17,-25){$\chi$}

\put(60,0){+}

\put(80,0){\line(1,0){60}}
\put(110,0){\circle{30}}
\put(107,20){$\phi$}
\put(107,3){{\scriptsize $\chi$}}
\put(107,-25){$\phi$}

\put(150,0){+}

\put(170,0){higher loop terms.}
\end{picture}
\label{selfchi}
\end{equation}

\vspace{1cm}

The setting sun (non-local) diagrams in (\ref{selfphi}) and 
(\ref{selfchi}) (the two last terms in (\ref{selfphi}) and 
(\ref{selfchi}))
contribute imaginary terms to the self-energies, from which we can  
write the decay
widths $\Gamma_\phi$, $\Gamma_{\chi_j}$, for the $\phi$ and $\chi_j$
fields, respectively, in terms of the on-shell expressions 
\cite{jeon2,heinz,parw} ($\Sigma_I \equiv {\rm Im} \Sigma$):

\begin{equation}
\Gamma_\phi (q) =  
\frac{\Sigma_I^\phi ({\bf q},\omega_\phi)}
{2 \omega_\phi}
\label{widthphi}
\end{equation}

\noindent
and

\begin{equation}
\Gamma_{\chi_j}(q) = \frac{\Sigma_I^{\chi_j}({\bf q},\omega_{\chi_j})}{2
\omega_{\chi_j}} \;,
\label{widthchi}
\end{equation}

\noindent
where $\omega_{\phi~(\chi_j)}$ is given by the solution of $\omega^2 =
{\bf q}^2 + m^2 + {\rm Re} \Sigma ({\bf q}, \omega)$.

Explicit expressions for $\Gamma (q)$ in the $\lambda \phi^4$ model
have been obtained in \cite{jeon2} and \cite{heinz}. We follow
\cite{jeon2} to compute $\Gamma_\phi$ and
$\Gamma_{\chi_j}$. A straightforward extension of the computation can
be applied to our model of interacting $\phi-\chi_j$ fields. Some of the
details are shown in Appendix A, where we evaluate
the imaginary contribution coming from the mixed setting sun diagrams 
in $\Sigma_\phi$ and $\Sigma_{\chi_j}$ 
[the last diagrams in (\ref{selfphi}) and (\ref{selfchi})]. Even though
in general there are no simple way of expressing the results, if we adopt
the zero space momentum ($|{\bf q}| =0$) approximation
for the imaginary part
of the self-energies, we can find simple approximated expressions for 
both (\ref{widthphi}) and (\ref{widthchi}),
respectively, given at finite temperature ($\beta=1/T$) 
by (for $m_T \sim {\cal O} (\mu_j (T))$)

\begin{eqnarray}
\Gamma_{\phi} (q)|_{\Sigma_I^\phi(0,m_T)} &\sim&
\frac{\lambda^2 T^2}{2^8 \pi^3 \omega_{\phi} ({\bf q})} \;
{\rm Li}_2  \left(1-e^{-\beta m_T} \right) \nonumber \\
&+& \left(1+\frac{1}{2} \delta_{\phi,\chi_j} \right)
\sum_{j=1}^N \frac{g_j^4 T^2}
{2^5 \pi^3 \omega_{\phi} ({\bf q})} \left[
{\rm Li}_2  \left(1-e^{-\beta m_T} \right) -
{\rm Li}_2 \left(\frac{1-e^{-\beta m_T}}{1-e^{-\beta \mu_j (T)}} \right)
\right]
\label{gammaphi}
\end{eqnarray}

\noindent
and 

\begin{eqnarray}
\Gamma_{\chi_j} (q)|_{\Sigma_I^{\chi_j}(0,\mu_j(T))} &\sim&
\frac{f_j^2 T^2}{2^8 \pi^3 \omega_{\chi_j} ({\bf q})} \;
{\rm Li}_2  \left(1-e^{-\beta \mu_j(T)} \right) \nonumber \\
&+& \left(1+\frac{1}{2} \delta_{\phi,\chi_j} \right)
\frac{g_j^4 T^2 }{2^5 \pi^3 \omega_{\chi_j} ({\bf q})} \left[
{\rm Li}_2  \left(1-e^{-\beta \mu_j(T)} \right) -
{\rm Li}_2 \left(\frac{1-e^{-\beta \mu_j(T)}}
{1-e^{-\beta m_T}} \right)
\right]
\: .
\label{gammachi}
\end{eqnarray}

\noindent 
In the above expressions, $m_T$ and $\mu_j(T)$ are the thermal
masses for $\phi$ and $\chi_j$, respectively. $\delta_{\phi,\chi_j} =1$
for $m_T=\mu_j(T)$ and $\delta_{\phi,\chi_j} =0$ otherwise.
${\rm Li}_2 (z)$ is the dilogarithm function\footnote{We follow the convention in 
Ref. \cite{stegun} for the 
definition of the dilogarithm function: 
${\rm Li}_2 (z)= - \int_1^z \frac{{\rm ln} t}{t-1} d t$. Some useful 
approximations for ${\rm Li}_2 (z)$ are ${\rm Li}_2 (z) \sim {{\pi^2}
\over 6} + 
[{\rm ln}(z)-1] z + 
{\cal O} (z^2)$, for $z \ll 1$, and 
${\rm Li}_2 (z) \sim - {{\pi^2}\over 6} - {1\over 2} \; {\rm ln}^2 (z) + 
{\cal O} (1/z)$, 
for $z \gg 1$.}.

This approximation for the decay widths, in terms of the zero 
space-momentum expression for the imaginary part of the self-energies, is
common 
to computations of transport coefficients and contrast densities
in field theory \cite{hs1,takao,jeon1}. However,  
Wang, Heinz and Zhang \cite{heinz} showed that this approximation may lead
to errors in the calculation of the contrast
density in the $\lambda \phi^4$ model.
In fact, the expressions for
${\rm Im} \Sigma$ can be fast changing for some momentum range and
values of the masses. {}For example, in Fig. 1 we plot the 
value of (the on-shell) ${\rm Im} \Sigma (q)$, 
obtained numerically, as a function
of the momentum, normalized
by its $|{\bf q}|=0$ expression (for $f_j \ll g_j^2$). Even though
${\rm Im} \Sigma (q)$ can depart considerable from its $|{\bf
q}|=0$ value, we will show later that, for a range of small thermal masses,
this approximation results in a small error
($\lesssim 10 \% $) in the expression for the dissipation coefficient, 
when compared with the computation using
the complete $|{\bf q}| \neq 0$
expressions for ${\rm Im} \Sigma (q)$.

In the analysis presented in the next sections, it will also be sufficient
to use the leading order high temperature expressions for the 
finite temperature effective (renormalized) masses, 
$m_T$ and $\mu_j (T)$, appearing in 
(\ref{widthphi}) - (\ref{gammachi}), 
(obtained from the 1-loop diagrams in (\ref{selfphi}) and (\ref{selfchi}),
respectively), given by\footnote{The divergences in (\ref{selfphi}) and 
(\ref{selfchi}), as in the effective action, can be dealt with by
the usual introduction of the appropriate renormalization counterterms 
in the initial Lagrangian, for the masses, coupling
constants and the wave-function. 
In particular, we note that the imaginary 
terms in the self-energies expressions, coming from the setting-sun 
diagrams, are finite. $m$, $\mu$, $g$, $f$ and $\lambda$ in (\ref{mT})
and (\ref{muT})
and in our later results are to be interpreted as the corrected and not
as bare quantities.}

\begin{equation}
m_T^2 = m^2 + {\rm Re}\Sigma_\phi (m_T)  \stackrel{T
\gg m}{\sim}  m^2 + \frac{\lambda T^2}{24}+  \sum_{j=1}^N g_j^2
\frac{T^2}{12} 
\label{mT}
\end{equation}

\noindent
and 

\begin{equation}
\mu_j^2 (T) = \mu_j^2 + {\rm Re}\Sigma_{\chi_j} (\mu_j
(T)) \stackrel{T \gg \mu_j}{\sim}  \mu_j^2 + \frac{f_j T^2}{24}+
g_j^2 \frac{T^2}{12} \:.
\label{muT}
\end{equation}

\subsection{Real-Time Full Field Propagators}

In order to obtain the evolution equation 
for the field configuration $\varphi$, we use
the real-time Schwinger's
closed-time path (CTP) formalism \cite{schw}. In the CTP formalism
the time integration is
along a contour $c$ from $- \infty$ to $+ \infty$ and then back
to $-\infty$. {}For reviews please see, for example,
refs. \cite{chou,rivers,weert}.

In the CTP formalism the field propagators are given by \cite{GR} 
(with analogous
expressions for $G_{\chi_j}$):

\begin{eqnarray}
\lefteqn{G_\phi^{++}(x,x') = i \langle T_{+} \phi(x) \phi(x')\rangle}
\nonumber \\
& & G_\phi^{--}(x,x') = i \langle T_{-} \phi(x) \phi(x')\rangle
\nonumber \\
& & G_\phi^{+-}(x,x') = i \langle \phi(x') \phi(x)\rangle
\nonumber \\
& & G_\phi^{-+}(x,x') = i \langle \phi(x) \phi(x')\rangle \: ,
\label{twopoint}
\end{eqnarray}

\noindent
where $T_{+}$ and $T_{-}$ indicate chronological and anti-chronological
ordering, respectively.
$G_\phi^{++}$ is the usual physical (causal)
propagator. The other three propagators come as a consequence of the
time contour and are considered as auxiliary (unphysical)
propagators. The expressions for
$G_\phi^{n,l}(x,x')$ in terms of its momentum-space Fourier
transforms are given by

\begin{equation}
G_\phi(x,x') = i \int \frac{d^3 q}{(2 \pi)^3} 
e^{i {\bf q} . ({\bf x} - {\bf x}')}
\left(
\begin{array}{ll}
G_\phi^{++}({\bf q}, t- t') & \:\: G_\phi^{+-}({\bf q}, t-t') \\
G_\phi^{-+}({\bf q}, t- t') & \:\: G_\phi^{--}({\bf q}, t-t')
\end{array}
\right) \: ,
\label{Gmatrix}
\end{equation}

\noindent
where

\begin{eqnarray}
\lefteqn{G_\phi^{++}({\bf q} , t-t') = G_\phi^{>}({\bf q},t-t')
\theta(t-t') + G_\phi^{<}({\bf q},t-t') \theta(t'-t)}
\nonumber \\
& & G_\phi^{--}({\bf q} , t-t') = G_\phi^{>}({\bf q},t-t')
\theta(t'-t) + G_\phi^{<}({\bf q},t-t') \theta(t-t')
\nonumber \\
& & G_\phi^{+-}({\bf q} , t-t') = G_\phi^{<}({\bf q},t-t')
\nonumber \\
& & G_\phi^{-+}({\bf q},t-t') = G_\phi^{>}({\bf q},t-t')
\label{G of k}
\end{eqnarray}

In terms of the decay width $\Gamma_\phi$, the expression 
for the full dressed propagators at finite temperature where obtained in
\cite{GR}, from which we have 

\begin{eqnarray}
& G_\phi^{>}({\bf q} ,t-t') &= \frac{1}{2 \omega_\phi}
\left\{ \left[ 1+ n(\omega_\phi - i \Gamma_\phi) \right]
e^{-i (\omega_\phi - i \Gamma_\phi) (t-t')} +
n(\omega_\phi + i \Gamma_\phi) e^{i (\omega_\phi +
i \Gamma_\phi) (t-t')} \right\}
\nonumber \\
& G_\phi^{<}({\bf q} , t-t') & = G_\phi^{>}({\bf q}, t'-t) \: ,
\label{G><}
\end{eqnarray}

\noindent
where $n(\omega)= \left(e^{\beta \omega} - 1\right)^{-1}$ is the Bose
distribution and $\omega\equiv \omega({\bf q})$ is the particle's energy,
or dispersion relation,
$\omega_\phi ({\bf q}) = \sqrt{{\bf q}^2 + m_T^2}$. {}For $G_{\chi_j}$,
$\omega_{\chi_j} ({\bf q}) = \sqrt{{\bf q}^2 + \mu_j^2 (T)}$.

\section{Dissipation in the Adiabatic Regime}

\subsection{The Effective Equation of Motion}

With fields in the forward and backward segments
of the CTP time contour identified as $\phi_+, \chi_+$ and
$\phi_-, \chi_-$,
respectively, the classical action can be written as

\begin{equation}
S[\phi,\chi] = \int d^4 x \left\{ {\cal L}[\phi_{+},\chi_+] - 
{\cal L}[\phi_{-},\chi_-]
\right\} \:,
\label{S+-}
\end{equation}

The evaluation of the effective action at real time can be done exactly
as in
\cite{GR}. There are also a number of other works using Schwinger's
closed-time path formalism to obtain the real-time effective action 
for field
configurations. [See. e.g., Refs. \cite{boya1,morikawa,greiner,yoko}.]
Here we will
concentrate on the evaluation of the effective equation of motion in the
strong dissipative regime. In the evaluation of the effective action
there appear several imaginary terms, once the $\chi$ fields and the 
fluctuations around the $\varphi$ background are integrated out. 
These imaginary terms can be interpreted as
coming from functional integrations over Gaussian stochastic fields, as
can be visualized by introducing the new field variables:

\begin{equation}
\varphi_c = \frac{1}{2} \left( \phi_+ + \phi_- \right)
\;\;,\;\;\; \varphi_\Delta = \phi_+ - \phi_- \;.
\end{equation}

In terms of these new variables the equation of motion is
obtained by \cite{GR,morikawa}

\begin{equation}
\frac{\delta S_{\rm eff}[\varphi_{\Delta},\varphi_c,\xi_j]}
{\delta \varphi_{\Delta}}|_{\varphi_{\Delta} = 0} = 0 ~~,
\label{motion}
\end{equation}

\noindent
where $\xi_j$ are stochastic fields, related to each
distinct dissipative kernel appearing in (\ref{motion}).

At 1-loop order, the leading contributions to the dissipative terms in
the equation of motion come from the diagrams:

\begin{equation}
\begin{picture}(270,28)

\put(50,0){\line(-1,-1){10}}
\put(60,0){\circle{20}}
\put(50,0){\line(-1,1){10}}
\put(70,0){\line(1,1){10}}
\put(70,0){\line(1,-1){10}}
\put(57,15){$\phi$}
\put(57,-20){$\phi$}

\put(88,0){+}
\put(100,0){$\sum_{j=1}^N \Big[$}

\put(150,0){\line(-1,-1){10}}
\put(160,0){\circle{20}}
\put(150,0){\line(-1,1){10}}
\put(170,0){\line(1,1){10}}
\put(170,0){\line(1,-1){10}}
\put(157,15){$\chi_j$}
\put(157,-20){$\chi_j$}

\put(190,0){$\Big]$}
\end{picture}
\label{leading}
\end{equation}

\vspace{0.5cm}

The explicit expression corresponding to these terms appearing in
the effective equation of motion, Eq. (\ref{motion}), is (as obtained in
\cite{GR} for a similar case)

\begin{eqnarray}
\lefteqn{ \varphi_c (x) \int d^4 x'
\varphi_c^2 (x') \left\{ \frac{\lambda^2}{2} 
{\rm Im}\left[G_\phi^{++}\right]_{x,x'}^2
+ \sum_{j=1}^N 2 g_j^4 
{\rm Im} \left[G_{\chi_j}^{++}\right]_{x,x'}^2 \right\} \theta(t-t')}
\nonumber \\
& & \sim \varphi_c^2 (t)
\dot{\varphi}_c (t) \; \left\{ \frac{\lambda^2}{8}
\beta \int \frac{d^3q}{(2 \pi)^3}
\frac{n_\phi(1 + n_\phi)}{\omega_\phi^2 ({\bf q}) \Gamma_\phi ({\bf q})} +
\sum_{j=1}^N \frac{g_j^4}{2} \beta \int \frac{d^3 q}{(2 \pi)^3}
\frac{n_{\chi_j} (1 + n_{\chi_j})}{\omega_{\chi_j}^2 ({\bf q})
\Gamma_{\chi_j} ({\bf q})} \right\}
 \nonumber\\
& & +{\cal O}\left(\lambda^2 \frac{\Gamma_\phi}{\omega_\phi}\right) +
{\cal O} \left(g_j^4 \frac{\Gamma_{\chi_j}}{\omega_{\chi_j}}
\right) \nonumber \\
& & +  \varphi_c^3 (t) \int_{-\infty}^t d t'
\int \frac{d^3 q}
{(2 \pi)^3} \left\{ \frac{\lambda^2}{2} 
{\rm Im} \left[G_\phi^{++}({\bf q},t-t')\right]^2  +
2 \sum_{j=1}^N g_j^4 {\rm Im} \left[G_{\chi_j}^{++}({\bf q},t-t')\right]^2
\right\}\: ,
\label{dissterm}
\end{eqnarray}

\noindent 
where in the lhs of the above equation, we used the compact notation

\begin{equation}
\left[G_{\phi,\chi_j}^{++}\right]_{x,x'}^2 = \int \frac{d^3 k}{(2 \pi)^3}
\exp\left [ i {\bf k} . ({\bf x} - {\bf x}')\right ] \int 
\frac{d^3 q}{(2 \pi)^3}
G_{\phi,\chi_j}^{++}({\bf q} ,t-t')G_{\phi,\chi_j}^{++}
({\bf q} - {\bf k},t-t') \;,
\label{G2}
\end{equation}

\noindent
with $G^{++}({\bf q},t-t')$
obtained from (\ref{G of k}) and (\ref{G><}). 
In the rhs of (\ref{dissterm}),
we have taken the limit of homogeneous fields, for details see the 
Appendix B.
We have also
made use of the
approximation for slowly moving fields: $\varphi_c^2 (t') -
\varphi_c^2(t) \sim 2 \varphi_c (t) \dot{\varphi_c} (t) (t'- t)$. In the
next section we show that this approximation is consistent with strong
dissipation. After performing the time integration and retaining the
leading terms in the coupling constants, we obtain the result given in
(\ref{dissterm}). The last term, proportional to $\varphi_c^3$, will
correspond to the finite temperature correction to the quartic $\phi$
self-interaction (see Appendix B).

The final equation of motion, at leading order in the coupling
constants, at high temperatures ($\mu_j (T), m_T \ll T$) and in the
adiabatic limit, can then be written as

\begin{equation}
\ddot{\varphi_c} +  m_T^2 \varphi_c (t) +
\frac{\lambda_T}{3 !} \varphi_c^3 (t) + \eta_1 \varphi_c^2 (t)
\dot{\varphi}_c (t) = \varphi_c (t) \xi_1 (t) \: ,
\label{eq motion}
\end{equation}

\noindent 
where $m_T$ is given by (\ref{mT}), $\lambda_T$ is the
temperature-dependent effective (renormalized) quartic coupling
constant\footnote{The terms linear in the temperature come from the
two-vertex diagrams in (\ref{leading}). The apparent
instability from these terms for high $T$ is only an artifact of the
loop expansion. As shown in \cite{fendley} for the $\phi^4$ model, once
higher order corrections are accounted for, $\lambda_T$ is always
positive even in the $T\to \infty$ limit. Using full dressed propagators
we are automatically taking into account these higher order corrections,
through the appearance of thermal masses in (\ref{lambdaT}). However, in
the multi-field case there is the possibility of vacuum instability due
to the $\phi$ couplings to the $\chi_j$ fields. This appears as a
constraint in our estimates below.}:

\begin{eqnarray}
\lambda_T  & \simeq & \lambda - \frac{3 \lambda^2}{2} \left\{
\frac{T}{8 \pi m_T} +
\frac{1}{8 \pi^2} \left[\ln\left(\frac{m_T}{4 \pi T} \right) +
\gamma \right]
+ {\cal O}\left(\frac{m}{T}\right) \right\} - \nonumber \\
& -& 6 \sum_{j=1}^N g_j^4 \left\{
\frac{T}{8 \pi \mu_j(T)} +
\frac{1}{8 \pi^2} \left[\ln\left(\frac{\mu_j(T)}{4 \pi T} \right) +
\gamma \right] + {\cal O}\left(\frac{\mu_j(T)}{T}\right) \right\}+
\nonumber \\
& + &
{\cal O}(\lambda^3, g^4 f, \lambda^2 g^2, g^6) \: ,
\label{lambdaT}
\end{eqnarray}

In (\ref{eq motion}), $\xi_1$ is a stochastic field associated with
the imaginary terms in the effective action coming from the real-time
evaluation of the diagrams (\ref{leading}). Its two-point correlation
function is given by \cite{GR}

\begin{equation}
\langle \xi_1 (x) \xi_1 (x')\rangle = \frac{\lambda^2}{2}
{\rm Re}\left[G_\phi^{++}\right]_{x,x'}^2  + 2 \sum_{j=1}^N
g_j^4 {\rm Re}\left[G_{\chi_j}^{++}\right]_{x,x'}^2\: .
\label{corr noise}
\end{equation}

\noindent
Note that since we are considering homogeneous field configurations,
$\xi_1$ is a space uncorrelated stochastic field, but it is colored
(time dependent) and Gaussian distributed, with probability distribution
given by ($N_1$ is a normalization constant)

\begin{equation}
P[\xi_1] = N_1^{-1} \exp\left\{ - \frac{1}{2} \int d^4 x d^4 x'
\xi_1 (x) \left[ \frac{\lambda^2}{2} {\rm Re}
\left[G_\phi^{++}\right]_{x,x'}^2
+ 2 \sum _j g_j^4 {\rm Re} \left[G_{\chi_j}^{++}\right]_{x,x'}^2
\right]^{-1} \xi_1 (x')
\right\} \: .
\label{P1}
\end{equation}

As shown in \cite{GR}, the dissipative coefficient in (\ref{eq motion}),
written explicitly in Eq. (\ref{disscoef}) below, and the noise
correlation function Eq. (\ref{corr noise}) (in the homogeneous limit),
are related by a fluctuation-dissipation expression valid within our
approximations (1-loop order at $\lambda^2, g_j^4$ and for
$\Gamma/\omega \ll 1 \: , \: \Gamma/T \ll 1$):

\begin{equation}
\eta_1 = \frac{1}{2 T} \int d^4 x' \langle \xi_1 (x) \xi_1 (x') \rangle
\:.
\label{fluc-diss}
\end{equation}

\noindent 
In \cite{GR} it was also shown that as $T \to \infty$,
$\Gamma_{\phi,\chi_j} \to \infty$, and the integrand in (\ref{corr
noise}) becomes sharply peaked at $|t-t^{\prime}|\sim 0$. In this limit,
we can obtain an approximate Markovian limit for (\ref{corr noise}).

We can read the dissipation coefficient $\eta_1$, which appears in Eq. 
(\ref{eq motion}), from (\ref{dissterm})\footnote{In \cite{GR} an extra
contribution to the $\phi$ decay rate coming from the $\phi-\chi$
interaction was left out. Here we give the correct expressions for
$\Gamma_\phi$, $\Gamma_\chi$ and for the dissipation.},

\begin{equation}
\eta_1 = \frac{\lambda^2}{8} \beta \int \frac{d^3 q}{(2 \pi)^3}
\frac{n_\phi (1 + n_\phi)}{\omega_\phi^2 ({\bf q}) \Gamma_\phi ({\bf q})}
+ \sum_{j=1}^N \frac{g_j^4}{2} \beta \int \frac{d^3 q}{(2 \pi)^3}
\frac{n_{\chi_j} (1 + n_{\chi_j})}{\omega_{\chi_j}^2 ({\bf q})
\Gamma_{\chi_j} ({\bf q})} +
{\cal O}\left(\lambda^2 \frac{\Gamma_\phi}{\omega_\phi}, \; 
g_j^4 \frac{\Gamma_{\chi_j}}{\omega_{\chi_j}} \right) \;.
\label{disscoef}
\end{equation}

{} For the model we are interested in, 
with Lagrangian density given by (\ref{Nfields}), with a large number of
$\chi$ fields coupled to $\phi$, and for 
$f_j \ll g_j^2$ and $\lambda \lesssim g_j$, we can use the obtained
expressions for $\Gamma_\phi$ and $\Gamma_{\chi_j}$, to show
that $\Gamma_\phi \gg \Gamma_{\chi_j}$. 
Since the dissipation coefficient, Eq. (\ref{disscoef}), 
goes as $1/\Gamma$, $\Gamma_{\chi_j}$
will give the dominant contribution to $\eta_1$.
An explicit expression for $\eta_1$, can be
obtained by using the $|{\bf q}|=0$ approximation for ${\rm Im} \Sigma_\phi
(q)$ and ${\rm Im} \Sigma_{\chi_j} (q)$, or, equivalently, Eqs.
(\ref{gammaphi}) and (\ref{gammachi}) for $\Gamma_\phi$ and
$\Gamma_{\chi_j}$, respectively, in Eq. (\ref{disscoef}). At the high
temperature limit, $T \gg m_T,\mu_j (T)$ 
and for $m_T \sim {\cal O} (\mu_j (T))$, 
we then obtain the following
approximate expression for $\eta_1$ (using ${\rm
Li}_2 (z) \sim \pi^2/6$, for $z \ll 1$)

\begin{eqnarray}
\eta_1 & \stackrel{T \gg m_T,\mu_j (T)}{\simeq} &
\frac{96}{\pi T} \left\{ \frac{\lambda^2}{8
\lambda^2 + \sum_{j=1}^N g_j^4 \left[1- \frac{6}{\pi^2} {\rm Li}_2 
\left(\frac{m_T}{\mu_j (T)} \right) \right] } 
\ln \left( \frac{2T}{m_T}\right)  \right. \nonumber \\
&+& \left. \sum_{j=1}^N \frac{4 g_j^4}{f_j^2 + 8 g_j^4 
\left[1- \frac{6}{\pi^2} 
{\rm Li}_2 
\left(\frac{\mu_j (T)}{m_T} \right) \right] }
\ln \left( \frac{2T}{\mu_j (T)}\right) \right\} \: .
\label{aproxdisscoef}
\end{eqnarray}

In order to test the validity of the above approximate expression for $\eta_1$,
we have computed it numerically.
The two expressions are shown in Fig. 2, for
$f_j \ll g_j^2$, $\lambda \lesssim g_j$, and $N=25$, where,
for simplicity, we have also considered $\mu_j = \mu$ and $g_j =g$ for
all $\chi_j$ fields ($m_T \sim 5 \mu_j (T)$). We see that the above
approximation for $\eta_1$ fits reasonably well the full expression for
the dissipation coefficient in the high temperature region, having a
$\lesssim 10 \%$ discrepancy for $m_T/T \lesssim
0.4$.

\subsection{Dissipation Coefficient and Shear Viscosity}

It is interesting to note the close relation of the above expression for
the dissipation coefficient with that obtained for the shear viscosity
evaluated, {\it e.g.}, from a Kubo formula \cite{takao,jeon1,jeon2}:

\begin{equation}
\eta_{\rm shear} = i \int d^3 {\bf x} \int_{-\infty}^0 dt \int_{-
\infty}^t dt' \langle [ \pi_{kl} (0), \pi_{kl} ({\bf x},t')] \rangle \;,
\label{shear1}
\end{equation}

\noindent
where $\pi_{kl} = (\delta^k_{\;i} \delta_{l}^{\;j} -
\frac{1}{3} \delta^k_{\;l} \delta_{i}^{\;j}) T^{i}_{\; j}$, with
$T^{i}_{\; j}$ the space components of the energy-momentum tensor. In
our case, with Lagrangian given by (\ref{Nfields}),

\begin{equation}
T^{\mu}_{\; \nu} = \frac{\partial {\cal L}}{\partial (\partial_\mu
\phi)} \partial_\nu \phi + \sum_j
\frac{\partial {\cal L}}{\partial (\partial_\mu
\chi_j)} \partial_\nu \chi_j - \delta^{\mu}_{\; \nu} {\cal L} \;.
\label{Tmunu}
\end{equation}

In order to compute the shear viscosity in (\ref{shear1}) to lowest
order, we must evaluate the diagrams (\ref{leading}), which, as shown in
\cite{jeon1,jeon2}, have near on-shell singularities coming from the
product of (bare) propagators. These singularities are softened once
explicit lifetimes for excitations are included through dressed
propagators. Taking this into account, we obtain the following
expression for the shear viscosity $\eta_{\rm shear}$ (in analogy with
the evaluation of $\eta_{\rm shear}$ in the $\lambda \phi^4$ single
field case),

\begin{equation}
\eta_{\rm shear} \stackrel{T \gg m_T,\mu_j (T)}{\simeq}
\frac{\beta}{30} \int \frac{d^3 k}{(2 \pi)^3} | {\bf k} |^4
\left[ \frac{n_\phi (1 + n_\phi)}{\omega_\phi^2  \Gamma_\phi } +
\sum_j \frac{n_{\chi_j} (1 + n_{\chi_j})}{\omega_{\chi_j}^2
\Gamma_{\chi_j} } \right] \;.
\label{shear2}
\end{equation}

\noindent 
Compare the above expression with (\ref{disscoef}). The
evaluation of (\ref{shear2}) leads to the standard result for the shear
viscosity being proportional to $T^3$ and inversely proportional to the
coupling constants. However, Eq. (\ref{shear2}), as shown by Jeon in
\cite{jeon2}, does not represent the unique contribution to $\eta_{\rm
shear}$ at this order of coupling constants. Due to the near on-shell
singularities and the way they are regulated by the thermal width, there
is an entire class of diagrams, called ladder diagrams (diagrams with
insertions of loops between the two propagators in (\ref{leading})),
contributing to $\eta_{\rm shear}$ at the same order. By using a formal
resummation of vertices, Jeon was able to perform the summation of the
whole set of ladder diagrams in the simple $\lambda \phi^4$ theory,
showing that the true value of the shear viscosity is about four times
larger than the one loop result in the high temperature limit. Since our
expression for the dissipation coefficient exhibits the same properties
of $\eta_{\rm shear}$, we expect that these higher loop ladder diagrams
will also give a significant contribution to the value of $\eta_1$ in
(\ref{disscoef}). However, as we are dealing with the more complicated
situation of several interacting fields, we will not attempt here to
evaluate these contributions. {}From the example of the shear viscosity
calculation in the single field case, these ladder contributions will
only add to the one-loop result for the dissipation coefficient, not
changing qualitatively our results. Thus, Eq. (\ref{disscoef})
represents, at least, a {\it lower bound} for the dissipation, applicable
in the strong dissipation regime, as we will show next.

\section{Adiabatic Approximation and Strong Dissipation}

We now investigate the validity and limits of applicability of our main
approximations, in particular the adiabatic approximation. In order to
arrive at the expression for the dissipation, Eq. (\ref{disscoef}), and
to write the equation of motion for $\varphi_c$ as in Eq. (\ref{eq
motion}), we assumed that the field $\varphi_c$ changes adiabatically
[see (\ref{dissterm})]:

\begin{equation}
\varphi_c^2 (t') -\varphi_c^2 (t) \simeq 2 (t'-t) \varphi_c (t)
\dot{\varphi_c} (t) + {\rm higher \; time\; derivative \; terms}\;.
\label{adiab}
\end{equation}

\noindent 
This approximation for the field configuration has recently
been the focus of some attention in the literature \cite{greiner}. The
authors in \cite{greiner}, working with soft field modes set by a coarse
graining scale $k_c$, showed that the adiabatic approximation breaks
down once the field configurations (soft modes) oscillate with the same
time scale as the dissipative kernels (with time scale given by $\sim
k_c^{-1}$). However, here we work in a very different context. We are
mainly concerned with the overdamped motion of the homogeneous field
configuration $\varphi_c$, {\it i.e.}, when its oscillatory motion is
suppressed. Therefore, the dynamic time-scale for $\varphi_c$ must be
much larger than the typical collision time-scale ($\sim \Gamma^{-1}$).
Note that this is a much stronger condition than the simple requirement
that the field should change slowly in time, with time scale set by the
frequency $\omega ({\bf k}) = \sqrt{{\bf k}^2 + m_T^2}$. Thus, we must
examine when the condition

\begin{equation}
\left| \frac{\varphi_c}{\dot{\varphi_c}} \right| \gg \Gamma^{-1}
\label{kinetic}
\end{equation}

\noindent
is satisfied.

We choose $\Gamma$ as the smallest of the two thermal
decay widths $\Gamma_\phi, \; \Gamma_{\chi_j}$, as it will set the
largest time-scale for collisions for the system in interaction with
the thermal bath.

Note that in the evaluation of the dissipation coefficient in
(\ref{dissterm}), the leading contribution to the first time derivative
of $\varphi_c$ is of order $\Gamma^{-1}$. As discussed earlier in
connection with the shear viscosity coefficient, the dependence of the
dissipation coefficient on the decay width $\Gamma$ comes from using it
as the regulator of on-shell singularities present in (\ref{leading}) at
first order in the time derivative. In Appendix B we present an argument
justifying the need of regulating with the decay width and also
compute the next order contribution in the adiabatic approximation,
showing the consistency of the results.

Since the stronger the dissipation the more efficient the adiabatic
approximation, the parameter range where (\ref{kinetic}) is valid leads
naturally to the regime where $\varphi_c$ undergoes overdamped motion
(in the sense of Eq. (\ref{over}) below).
If we consider the ensemble average of the equation of motion (\ref{eq
motion}):

\begin{equation}
\left\langle \frac{\delta S_{\rm eff}[\varphi_{\Delta},\varphi_c,\xi_j]}
{\delta \varphi_{\Delta}}|_{\varphi_{\Delta} = 0} \right\rangle = 0 \;,
\label{aver}
\end{equation}

\noindent
where $\langle \ldots \rangle$ means average over the stochastic fields,
then we define the overdamped regime when the (averaged) background
configuration $\varphi_c$ satisfies

\begin{equation}
\eta_1 \varphi_c^2 \dot{\varphi_c} + m^2_T \varphi_c +
\frac{\lambda_T}{6} \varphi_c^3 = 0 \;.
\label{over}
\end{equation}

We also restrict our study to the high-temperature and
ultra-relativistic region: $T \sim |{{\bf q}}| \gg m_T,\mu_j (T)$. We
take the couplings $g_j$, $f_j$ such that $g_j^2 \gg f_j$. Also, for
simplicity, as before, we take all $g_j = g$. At high temperatures we
can then write for (\ref{mT}) and (\ref{muT}) ($T^2 \gtrsim 24
m^2/\lambda, \; 12 \mu^2/g^2$) the expressions

\begin{equation}
m_T^2 \sim \left( \lambda + 2 N g^2 \right) \frac{T^2}{24}\;,
\label{newm}
\end{equation}

\noindent
where $N$ is the number of fields coupled to $\phi$, and

\begin{equation}
\mu_T^2 \sim g^2 \frac{T^2}{12}~.
\label{newmu}
\end{equation}

\subsection{Results for three different cases}

We will examine the condition for strong dissipation with overdamped
motion for three particular choices of parameters, showing that there is
a region of parameter space consistent with this regime. Using
(\ref{over}), we can write the equivalent expression for
(\ref{kinetic}):

\begin{equation}
\left| \frac{m_T^2 + \frac{\lambda_T}{6} \varphi_c^2}{\eta_1 \varphi_c^2} 
\right| \ll \Gamma_\chi \;.
\label{kinetic2}
\end{equation}

\noindent
In the estimates below, we evaluated both 
$\eta_1$ [from Eq. (\ref{disscoef})], and $\Gamma_{\chi}$ 
(computed at $|{\bf q}| = T$) numerically.
The three cases analyzed are:

\noindent
{\bf Case 1:} $\lambda \sim g^2$: In this case we obtain that

\begin{equation}
m_T^2 \sim (1+2N) g^2 \frac{T^2}{24}~,
\end{equation}

\begin{equation}
\lambda_T~ \gtrsim ~g^2\left(1 - \frac{3 \sqrt{3} N g}{2 \pi}\right) ~.
\end{equation}

Note that the last condition is written as a constraint for the positivity
of $\lambda_T$. With these values and for the case $N=25$,
we obtain the results shown in Fig. 3a, where we have plotted
both sides of Eq. (\ref{kinetic2}). 
The region of parameters satisfying
Eq. (\ref{kinetic2}) is given by the intersection of the region 
below the solid lines (the function $\Gamma_\chi$) with the region
above the dashed line ($|\dot{\varphi}_c/\varphi_c|$ computed for
different values of $\varphi_c$). 

\noindent
{\bf Case 2:} $\lambda \sim g$: As above, this is shown in Fig. 3b. The
region satisfying Eq. (\ref{kinetic2}) is given again by the intersecting
region below the  solid line and  above the dashed lines. 

In both Figs. 3a and 3b, the results are shown up
to the value of $m_T$ satisfying the constraint for the positivity
of $\lambda_T$.

\noindent
{\bf Case 3:} $\lambda_T\approx g^4$: This case follows a 
slightly different 
philosophy, of fixing the corrected
coupling as opposed to the bare coupling. We have,

\begin{equation}
\label{3lam}
\lambda \approx g^4 +
\frac{3 \sqrt{3} N g^3}{2 \pi}~\equiv \lambda(g,N)~,
\end{equation}

\begin{equation}
\label{3mass}
m_T^2 \sim \left(\lambda(g,N) + 2 N g^2\right)
\frac{T^2}{24}~,
\end{equation}

\noindent
with the additional constraint,

\begin{equation}
\lambda(g,N) < 1~.
\label{lambcond}
\end{equation}

\noindent 
The results for this case are shown in Fig. 3c, with the same 
interpretation as
in cases 1 and 2: the region satisfying Eq. (\ref{kinetic2}) is given
by the intersecting region below the solid line and above the
dashed lines. The results are shown up to the value for $m_T$ satisfying
the condition (\ref{lambcond}). 

We note that the case $\lambda_T \approx g^4$ is the one with the 
broadest range of validity in parameter space, as seen in Fig. 3c, 
followed by the case $\lambda \approx g$, shown in Fig. 3b. {}For
$\lambda = g^2$, the condition for adiabaticity is only possible
for fairly large field amplitudes, which may be beyond the validity of
a perturbative evaluation of the effective action. We will come back to
this issue in the next section. In any case, we stress that there are
several regimes where the adiabatic approximation is valid. 

In all cases, the smaller $N$ the
smaller the region of parameters that satisfies (\ref{kinetic}). In
particular, for $N<2$, we find no parameter range satisfying
(\ref{kinetic}) and therefore, the adiabatic approximation. This is
consistent with the intuition that dissipation is caused by the decay of
the $\phi$ field into $\chi$ fields and is more efficient the larger the
number of decay channels available. We also obtain that $\varphi_c$ is
always somewhat large ( $\gtrsim 2 T$) for the range of physical 
parameters
satisfying (\ref{kinetic}), for both cases analyzed, being even higher
for case 1. 

If in (\ref{kinetic2}) we use $\Gamma_\phi$ instead of
$\Gamma_{\chi_j}$, the region of parameters improves considerably; since
$\Gamma_\phi \gg \Gamma_{\chi_j}$ for large $N$, it allows 
much smaller values of $\varphi_c/T$. It should be recalled that
$\Gamma_\phi$ determines the relaxation time scale for the $\phi$ field.

{}Finally, as discussed earlier, the expression we quoted for $\eta_1$
gives only a lower bound for the dissipation coefficient. 
As in the case studied by Jeon in \cite{jeon2}, higher loop ladder
diagrams can lead to a considerably higher value for $\eta_1$.
For several interacting fields, simple estimates show that
these ladder diagrams scale at most as $N$.
Therefore, they may well be of the same order as
the leading order 1-loop contribution to the dissipation coefficient,
given by the $\chi$-sector. We leave a more detailed analysis of
the contributions coming from ladder diagrams to a future work.
Additional contributions to the dissipation coefficient in
(\ref{disscoef}) only improve our estimates, enlarging the region of
parameter space satisfying the adiabatic approximation; 
the ratio $\varphi_c/T$ decreases, broadening 
the conditions under which the field
undergoes overdamped motion (strong dissipative regime).

It is worth mentioning that the $\phi-\chi$ coupling constant 
in Eq. (\ref{Nfields})
can be negative and this also leads to interesting results. As an
illustrative example, consider an even number of $\chi$-fields with the
sign of the $\phi-\chi$ coupling distributed so that
\begin{equation}
V_{\rm int} = \sum_{j=1}^{2k} (-1)^j g^2 \phi^2 \chi_j^2
\end{equation}
and $f_j=f, j= 1 \ldots 2k$.  In order that the potential be strictly
positive, it requires 
\begin{equation}
\left( \frac{\lambda}{24} + \frac{Nf}{24} - \frac{Ng^2}{4} \right) > 0,
\end{equation}
which for large $N$ implies $g^2 < f/6$. 
In the alternating sign regime (ASR) the thermal masses are
\begin{equation}
m_T^2 \approx \frac{\lambda}{24} T^2
\end{equation}
and
\begin{equation}
\mu_T^2 \approx \frac{f}{24} T^2.
\end{equation}
Following an analysis similar to above and for case 3 ($\lambda_T\sim g^4$), 
we find a
solution regime within the perturbative amplitude expansion,
$g \phi < \mu_T$, $\lambda \phi < T$, for 
$g^2 < f^{3/2} {\rm ln}(2\sqrt{24/f})/46$ 
and $N \sim 1/g^4$.
For example, these conditions are satisfied for $f \lesssim 1.0$,
$g^2 \lesssim 1/20$.  In this example $\lambda  \sim g^2$, but this 
can be modified in several ways.
In general, when
the $\phi-\chi$ couplings are distributed between positive and negative
strengths, it controls the growth of $m_T$ 
due to the cancellation of thermal mass
contributions from the $\chi$-fields.  
Restricting the magnitude of $m_T$, in turn, increases the 
parameter regime and duration
of overdamped motion.   This example demonstrates another
regime of overdamped motion
in our model for small field amplitudes 
$ g^2 \varphi_c^2 < \mu_T^2$.

\subsection{Summing up the whole 1-loop series - The effective potential}

The fact that overdamping in (\ref{eq motion}) for much of the parameter
space demands large field amplitudes, at least within the approximation
scheme used here, is a direct consequence of having a field dependent
dissipation $\eta(\varphi) \sim \varphi^2$. Since in (\ref{graphs1l}) we
are considering a perturbative expansion for the 1-loop effective action
in the field amplitudes (that is, in powers of $\lambda \varphi_c^2/2$
and $g_j^2 \varphi_c^2$), the need for large field amplitudes may place
doubts on the validity of our calculations for a considerable portion of
the parameter space. Below we address this issue in two different ways; first
by comparing our results with an improved one-loop approximation and then by
using the subcritical bubbles method \cite{subcrit} to test the validity of the
effective potential for large-amplitude fluctuations. 

We start by computing the analog
of (\ref{over}) in the context of the whole one-loop approximation, {\it
i.e.}, when $\lambda \varphi_c^2/2$ and $g_j^2 \varphi_c^2$ in
(\ref{Trln}) are taken as part of field-dependent masses. {}For this,
let us give an alternative computation of the evolution equation for
$\varphi_c$ in terms of the tadpole method of Weinberg
\cite{boya1,weiss,ringwald}: in the shifting of the scalar field, $\phi
= \varphi_c + \eta$, the requirement $\langle \eta \rangle=0$ leads, at
the one-loop order, to the evolution equation for $\varphi_c$ (for
homogeneous fields)

\begin{equation}
\ddot{\varphi}_c + m^2 \varphi_c + \frac{\lambda}{6} \varphi_c^3 +
\frac{\lambda}{2} \varphi_c \langle \eta^2 \rangle + \sum_j g_j^2 
\varphi_c \langle \chi_j^2 \rangle = 0 \;,
\label{eqnew}
\end{equation}

\noindent
where $\langle \eta^2 \rangle$ and $\langle \chi_j^2 \rangle$  are
given in terms of the coincidence limit of the (causal)
two-point Green's functions $G^{++}_\phi (x,x')$ and $G^{++}_{\chi_j}
(x,x')$, respectively, which satisfy, in the fully dressed propagator
form (see, e.g., Ringwald in \cite{ringwald})

\begin{equation}
\left[\Box + m^2 + \frac{\lambda}{2} \varphi_c^2 \right] G^{++}_\phi
(x,x') + \int d^4 z \Sigma_\phi (x,z) G^{++}_\phi (z,x') = i \delta
(x,x')
\label{Gphi}
\end{equation}

\noindent
and 

\begin{equation}
\left[ \Box + \mu_j^2 + g_j^2 \varphi_c^2 \right] G_{\chi_j}^{++} (x,x')
+ \int d^4 z \Sigma_{\chi_j} (x,z) G^{++}_{\chi_j} (z,x') = i \delta
(x,x')\;,
\label{Gchi}
\end{equation}

\noindent
where, in (\ref{Gphi}) and (\ref{Gchi}), $\Sigma_{\phi} (x,x')$ and 
$\Sigma_{\chi_j} (x,x')$
are the (causal) self-energies for the $\phi$ and $\chi_j$ fields,
respectively. By expressing $\eta (x)$ and $\chi_j (x)$ in terms of 
mode functions, we can then evaluate the averages in (\ref{eqnew}).
An explicit expression can be obtained in the approximation (equivalent to
the adiabatic approximation)
$\dot{\omega}_\phi (\varphi_c)/\omega_\phi^2 (\varphi_c) \ll 1$
and $\dot{\omega}_\chi (\varphi_c)/\omega_\chi^2 (\varphi_c) \ll 1$, for
which there is a WKB solution for the mode functions of the fields.
In this paper, however, we will not carry out this calculation. 
A detailed study of this, in the context of an expanding background and
along the proposals made in the next section, will be presented in a 
forthcoming paper. 

{}For now, we can present the result of this
calculation, by using the simplest formulation proposed in \cite{hs1}, 
based on a relaxation-time approximation of the kinetic equation, 
for the calculation of the averages in (\ref{eqnew}). We can 
then show that the (ensemble averaged) evolution equation for 
$\varphi_c (t)$ can be expressed, in the quasi-adiabatic approximation
(hydrodynamical regime of \cite{hs1}), by

\begin{equation}
\ddot{\varphi}_c + V^\prime_{\rm eff} (\varphi_c)
+ \eta_1 \varphi_c^2 \dot{\varphi}_c = 0 \;,
\label{eqVeff}
\end{equation}

\noindent
where $V^\prime_{\rm eff} (\varphi_c)= \frac{\partial 
V_{\rm eff} (\varphi_c)}{\partial \varphi_c}$, is the field derivative
of the 1-loop effective potential, 

\begin{eqnarray}
V^\prime_{\rm eff} (\varphi_c) & = &
m^2 \varphi_c + \frac{\lambda}{6} \varphi_c^3 + \frac{\lambda}{4}
\varphi_c \int \frac{d^3 q}{(2 \pi)^3} \frac{1 + 2 n
(\omega_\phi)}{\omega_\phi}  \nonumber \\
& + & \sum _j \frac{g_j^2}{2}
\varphi_c \int \frac{d^3 q}{(2 \pi)^3} \frac{1 + 2 n
(\omega_{\chi_j})}{\omega_{\chi_j}} \;,
\label{Veff}
\end{eqnarray}

\noindent 
where $\omega^2_\phi = {\bf q}^2 + m_T^2 + \frac{\lambda}{2}
\varphi_c^2$ and $\omega^2_{\chi_j} = {\bf q}^2 + \mu_j^2 (T) + g_j^2
\varphi_c^2$ are the field dependent frequencies, with masses given
in terms of the thermal ones\footnote{Note that this will lead to the daisy 
corrected effective potential. In particular, once the thermal masses are
being introduced in the derivative of the effective potential, expressed
in terms of one-loop tadpole graphs in the Weinberg method, it is well
known that this method leads to a consistent finite temperature effective 
potential \cite{lindeew}, with daisy graphs incorporated.}, 
Eqs. (\ref{mT}) and (\ref{muT}). Also, $\eta_1$ in (\ref{eqVeff}) is the
same as in (\ref{disscoef}), but now with the masses replaced by the field
dependent ones.  

In terms of (\ref{eqVeff}), in the overdamping approximation, the
condition (\ref{kinetic}) becomes

\begin{equation}
\left| \frac{V_{\rm eff}^\prime}{\eta_1 \varphi_c^3} \right| \ll \Gamma
\;.
\label{kin2}
\end{equation}

\noindent 
Using Eq. (\ref{kin2}), in the high temperature approximation
for the fields, $m_\phi (T)/T, m_{\chi_j} (T)/T \ll 1$, we can show that
the results obtained earlier, in terms of the amplitude expansion for the
effective action, for instance, the results expressed in Fig. 3 (with
$m_T$ replaced with the field dependent mass $m_\phi (T)$), remain
approximately the same, for the cases where $\varphi_c \lesssim 2 T$. Thus,
at least for these values of the field amplitude, higher order corrections do 
not add to the effective potential. In other words,
at leading order in the
high-temperature expansion, the field derivative of $V_{\rm eff}$ can be
just expressed as in (\ref{over}), $V_{\rm eff}^\prime \sim m_T^2
\varphi_c + \lambda_T/6 \, \varphi_c^3$. 

We can also address the issue of high-amplitude fluctuations by adopting
a method suggested in Ref. \cite{G+R}, where it was applied to test the
validity of the 1-loop approximation to the electroweak effective potential.
We note that the results from this approach are entirely consistent with
nonperturbative computations based on lattice gauge theories
performed by K. Kajantie {\it et al.} \cite{Kajantie}.

The interactions of the field $\varphi$ with a thermal environment will
promote fluctuations around the perturbative vacuum. The subcritical
bubbles method models these fluctuations as unstable spherically-symmetric
configurations with a distribution of sizes and amplitudes. For details
see Refs. \cite{subcrit,G+R}. Using a distribution function for these
configurations, it is possible to compute the RMS amplitude of the 
fluctuations \cite{G+R},

\begin{equation}
{\bar \varphi}(T) = \sqrt{\langle \varphi^2\rangle_T-
\langle \varphi\rangle_T^2}~,
\end{equation}

\noindent
where $\langle \dots\rangle$ is the thermal average defined in Ref. \cite{G+R}.
{}For the effective potential of Eq. (\ref{over}), we obtain,

\begin{equation}
{\bar \varphi}^2(T) \simeq {1\over 6}m_TT~.
\end{equation}

Since the perturbative approach for the 
computation of the effective potential relies on a saddle-point approximation
to the partition function, it will only be valid for small-amplitude
fluctuations about the perturbative vacuum. 
{}For potentials which exhibit spontaneous symmetry breaking,
it is customary to choose the maximum amplitude to be at the inflection
point, $\varphi_{\rm max} \lesssim \varphi_{\rm inf}$. Here, since we have a
potential with positive-definite curvature, we will
conservatively assume that the perturbative expansion is valid for fluctuations
dominated by the quadratic term of the effective potential, that is, for

\begin{equation}
\varphi_{\rm max}^2 \lesssim 12{{m_T^2}\over {\lambda_T}}~.
\end{equation}

The condition for the validity of the 1-loop approximation for the effective
potential is then written as

\begin{equation}
{\bar \varphi}^2(T) \leq \varphi_{\rm max}^2~.
\label{subcond}
\end{equation}

It is straightforward to apply this condition to the 3 cases analysed above.
Since case 3 is the one with a larger range of parameters satisfying
the adiabatic condition, we use it as an illustration.
{}From Eqs. (\ref{3lam}) and (\ref{3mass}), we can write Eq. (\ref{subcond}),
after dividing by $T^2$, as the inequality

\begin{equation}
[f(g,N)]^{1/2} > {{g^4}\over {144\sqrt{6}}}~,
\end{equation}

\noindent
where, $f(g,N)\equiv 24m_T^2/T^2$. This condition is easily satisfied
for a large range of parameters. In particular, for 
$\lambda=0.5,~g=0.3,~N=25$, which are values inside the region of 
parameters allowed for 
overdamping shown in Fig. 3c for $\varphi_c \gtrsim 2 T$,
we obtain, ${\bar \varphi}\simeq 0.3T$ and $\varphi_{\rm max}\simeq 16T$,
well within the range of validity of the small-amplitude approximation.
We thus conclude that it is possible to attain the adiabatic limit of
strong dissipation within the 1-loop approximation scheme adopted here.

\section{Applying Strong Dissipation to Warm Inflation}
\label{sec5}

The calculation in Secs. II-IV presented a microscopic quantum field
theory model of strong dissipation in Minkowski spacetime. This section
addresses the application of this calculation to the cosmological warm
inflation scenario. Although we will not present a detailed extension of
our previous results to an expanding spacetime, we will argue that most
of the modifications are quite straight-forward up to the requirements
for the warm inflation scenario.

\subsection{Formulation}

Consider the standard
Friedmann cosmology with Robertson-Walker metric

\begin{equation}
ds^2 = dt^2 - R^2(t)\left[\frac{dr^2}{1-kr^2}
+r^2d\theta^2+r^2 \sin^2\theta d\phi^2 \right].
\end{equation}

\noindent 
We restrict our analysis to flat space, $k=0$, and quasistatic
de Sitter expansion, $H \equiv {\dot R}/R \approx {\rm const.}$ For
notational convenience, the origin of cosmic time is defined as the
beginning of our treatment. {}For this metric, the 
minimally coupled Lagrangian for the
model in Eq. (\ref{Nfields}) is

\begin{eqnarray}
\lefteqn{L = \int_V d^3{\bf x} e^{3Ht}\left\{ \frac{1}{2} \left(
(\partial_0 \phi({\bf x},t))^2-
(e^{-Ht}{\bf \nabla} \phi({\bf x},t))^2 - m^2 \phi^2({\bf x},t) \right)
-V(\phi({\bf x},t)) \right.}
\nonumber\\
& & +\left. \sum_i  \frac{1}{2}
\left[ (\partial_0 \chi_i({\bf x},t))^2-
(e^{-Ht}{\bf \nabla} \chi_i({\bf x},t))^2 - \mu_i^2 \chi_i^2({\bf x},t)
- \frac{f_i}{12} \chi^4_i({\bf x},t)
\right]  \right. \nonumber \\
& & - \left. \sum_i \frac{g_i^2}{2}
\chi^2_i({\bf x},t) \phi^2({\bf x},t) \right\}.
\label{langran}
\end{eqnarray}

There exists an alternative derivation of the ensemble average of Eq.
(\ref{eq motion}), which was presented for a single scalar field, as an
intuitive argument in \cite{hs1}. In the context of a single scalar
field, the method is to work directly with the operator equation of
motion for $\phi({\bf x},t)$. The operator $\phi$ is reexpressed as the
sum of a c-number $\varphi_c(t)$, representing the classical
displacement, plus a shifted operator $\eta({\bf x},t)$,
\begin{equation} \phi({\bf x},t) =\varphi_c(t)+\eta({\bf x},t)
\label{phiceta} \end{equation} with $\langle \phi({\bf x},t)
\rangle_{\beta}= \varphi_c(t)$. A thermal average is taken of this
equation of motion, in which thermal expectation values involving
$\eta({\bf x},t)$ are computed such that $\varphi_c(t)$ is treated as an
adiabatic parameter. To the order of perturbation theory considered in
the previous sections, for the single scalar field the intuitive
derivation in \cite{hs1} gives the same effective equation of motion as
the ensemble average of Eq. (\ref{eq motion}) as shown in \cite{GR}.

No new considerations are needed to apply this intuitive derivation to
the model in Eq. (\ref{Nfields}). The $\chi_i({\bf x},t)$ fields are
treated as quantum fluctuations similar to $\eta({\bf x},t)$. From the
treatment in \cite{hs1}, it follows that the expressions for
$m_T,\lambda_T$ and $\eta_1$ in Eq. (\ref{eq motion}) will arise from
the thermal averages, $\langle \eta^2({\bf x},t) \rangle_{\beta}$ and
$\langle \chi_i^2({\bf x},t) \rangle_{\beta}$, taken with respect to the
instantaneous background $\varphi_c(t)$.

Although the approach in \cite{hs1} immediately isolates the dissipative
term and the finite temperature renormalizations at the level of the
equation of motion, it is not systematic to all orders. Furthermore, it
cannot treat noise and it is valid only in the adiabatic approximation.
These limitations can be accounted for in the closed-time-path formalism
used in this paper. A recent work \cite{boya2} has discussed some of the
difficulties associated with extending this formalism to an expanding
background in order to treat noise and dissipation. Our goal at present
is more modest. As an easier first step, the intuitive derivation of
\cite{hs1} is extended to an expanding background.

The exact operator equations of motion from the Lagrangian
Eq. (\ref{langran}) are

\begin{equation}
\ddot{\phi} ({\bf x}, t) +3H{\dot \phi}({\bf x},t)
-e^{-2Ht}\nabla^2 \phi({\bf x},t)+
m^2 \phi({\bf x},t)+
\frac{\delta V}{\delta \phi({\bf x},t)}
+\sum_i g_i^2 \phi({\bf x},t) \chi^2({\bf x},t)=0 \;,
\label{phieq}
\end{equation}

\noindent
and

\begin{equation}
\ddot{\chi}_i ({\bf x}, t) +3H{\dot \chi}_i({\bf x},t)
-e^{-2Ht}\nabla^2 \chi_i({\bf x},t)
+ g_i^2 \chi_i({\bf x}, t) \phi^2({\bf x},t)+ \mu^2_i\chi_i({\bf x}, t)
+ {{f_i}\over 6}\chi_i({\bf x}, t)^3=0 \;.
\label{chieq}
\end{equation}

\noindent The objective is to displace the operator $\phi({\bf x},t)$ by
a x-independent c-number at time $t=0$, $\langle \phi({\bf x},t=0)
\rangle_{\beta} = \varphi_c(0)$, and then determine the evolution of the
expectation value $\langle \phi({\bf x},t) \rangle_{\beta} \equiv
\varphi_c(t)$ by solving Eqs. (\ref{phieq}) and (\ref{chieq})
perturbatively. Thus $\phi({\bf x},t)$ is reexpressed as Eq.
(\ref{phiceta}). With this definition of $\varphi_c(t)$, for flat,
$k=0$, nonexpanding, $H=0$, spacetime, the resulting equation of motion
is the same as the ensemble average of the equation of motion, Eq.
(\ref{eq motion}).

{}For the case of expanding spacetime, $H \ne 0$, in order to obtain the
equation of motion for $\varphi_c(t)$, thermal expectation values must
be taken of Eqs. (\ref{phieq}) and (\ref{chieq}). Provided the
temperature, $1/\beta$, of the thermal bath is time independent, ({\it
i.e.}, rapid equilibration time scales), thermal expectation values of
terms linear in $\phi({\bf x},t)$ can be replaced by $\varphi_c(t)$,
just as for the nonexpanding case. In evaluating $\langle \eta^2({\bf
x},t) \rangle_{\beta}$ and $\langle \chi_i^2({\bf x},t)
\rangle_{\beta}$, if the characteristic time scale for the quantum
fluctuations is much faster than the expansion time scale, $1/H$, the
calculation is no different from the Minkowski space situation. This
criteria is self-consistently satisfied provided 

\begin{equation}
\Gamma_{\chi}, \Gamma_{\phi} \gg H \;, 
\label{ggth} 
\end{equation}

\noindent 
where the left hand side is given in Eqs. (\ref{gammaphi}) and
(\ref{gammachi}) for our model.

These arguments suggest that at leading nontrivial order, the effective
equation of motion for $\varphi_c(t)$ in an expanding de Sitter spacetime,
under the same conditions required for Eq. (\ref{eq motion}) plus the
additional condition Eq. (\ref{ggth}) is

\begin{equation}
\ddot{\varphi}_c (t)+[\eta_1 \varphi^2_c(t)+3H]{\dot \varphi}_c(t)
+m_T^2 \varphi_c(t)+ \frac{\lambda_T}{6} \varphi^3_c(t)=0.
\label{phiheq}
\end{equation}

\noindent 
{}Further justification that Eq. (\ref{phiheq}) is the
appropriate replacement of Eq. (\ref{eq motion}), for expanding de
Sitter space, can be obtained from \cite{ab54}, where an effective
equation of motion similar to Eq. (\ref{phiheq}) was obtained for a
model like Eq. (\ref{langran}). However, the coupling between fields was
linear, $\phi \chi_i$, which is analytically much more tractable than
the present case of quadratic coupling, $\phi^2 \chi^2_i$.

The entire discussion above assumes that the temperature has a well
defined meaning in an expanding background. Furthermore, Eq.
(\ref{phiheq}) has been motivated under the restriction Eq.
(\ref{ggth}). As will be discussed next, condition (\ref{ggth}) is a
specific example of a general microscopic property argued in \cite{ab54}
to be a necessary condition for warm inflation. As such, when
(\ref{phiheq}) is applied to the warm inflation scenario, condition
(\ref{ggth}) imposes no additional restriction.

The warm inflation picture requires that an order parameter, in a strongly
dissipative regime, slowly rolls down a potential, liberating vacuum
energy into radiation energy, $\rho_r$. The nonisentropic expansion
which underlies warm inflation imposes that the rate of radiation
production is sufficient to compensate for red-shift losses due to
cosmological expansion, 

\begin{equation} H \gg
\frac{|\dot{\rho}_r|}{\rho_r}. 
\label{rholth1} 
\end{equation} 

\noindent
To give
meaning to temperature, the newly liberated radiation must thermalize at
a scale $\Gamma_{\rm rad}$ which is faster than the expansion scale,

\begin{equation} 
\Gamma_{\rm rad} \gg H. 
\label{rholth} 
\end{equation}

\noindent 
Minimally this requires an energy transfer rate from vacuum to
radiation that is faster than the expansion rate, which in our model
implies the condition (\ref{ggth}). Thus, Eq. (\ref{rholth}) is
necessary to justify a temperature parameter $T$, which, combined with
condition (\ref{ggth}), are sufficient to justify the arguments leading
to Eq. (\ref{phiheq}). To completely justify a temperature parameter for
an expanding background spacetime, it is required studying the
thermalization of the radiation, once it is liberated. General
arguments, as well as specific calculations \cite{ellstg,enq} at high
temperature, indicate that this rate is set by the temperature
$\Gamma_{\rm rad} \sim \alpha T$, for some appropriate, model dependent,
coefficient $\alpha$. This minimally requires $T \gg H$. However,
$\alpha$ may be very small, as for example in Eqs. (\ref{gammaphi}) and
(\ref{gammachi}). Thus the correct constraint is

\begin{equation} 
\alpha T \gg H. 
\label{hllat} 
\end{equation} 

\noindent
This
problem will not be considered further here. Eq. (\ref{ggth}) will be
our only criteria for thermalization. This is equivalent to assuming
that the thermalization rate is at least as fast as the energy transfer
rate.

Once Eq. (\ref{phiheq}) is accepted as the macroscopic equation
governing the evolution of the order parameter $\varphi_c(t)$, it can be
used as a given input to construct warm inflation scenarios as in
\cite{wi,ab55}. The microscopic origin of the equation can be forgotten
up to restrictions on parameters and the self consistency condition Eq.
(\ref{ggth}). For a general equation like Eq. (\ref{phiheq}), the warm
inflation scenario requires the strong dissipative regime \cite{wi}:

\begin{equation}
\left[ \eta(\varphi_c) +3H \right] {\dot \varphi}_c 
\gg {\ddot \varphi}_c \;,
\label{ddpledp}
\end{equation}

\noindent 
with $\eta(\varphi_c)=\eta_1 \varphi^2_c$ for our model. For the
derivation in Secs. II-IV, where $H=0$, this condition is sufficient to
satisfy the adiabatic condition, Eq. (\ref{kinetic}), which is required
for the consistency of the microscopic calculation. As such, this model
provides an example of a general point conveyed in \cite{ab54}, that
warm inflation defines a good regime for application of finite
temperature dissipative quantum field theory methods. The study of warm
inflation in \cite{ab54,wi,ab55} also found that to satisfy
observational constraints on the expansion factor, it requires 

\begin{equation} 
\eta(\varphi_c) \gg 3H. 
\label{eggh} 
\end{equation} 

\noindent 
Thus, warm inflation is an extreme example of dissipative dynamic
during de Sitter expansion. As demonstrated in \cite{bf2,rudnei},
dissipation is generally prevalent during inflation. The microscopic
model in this paper could be used to examine the general case, but then
the condition (\ref{eggh}) can be relaxed. Here, only the warm
inflation regime will be further examined. Thus in the limit given by
Eq. (\ref{eggh}) and based on the remaining discussion in this section,
the equation of motion for the order parameter $\varphi_c(t)$ in our
model for the warm inflation scenario turns out to be Eq. (\ref{over})
but with the additional constraint Eq. (\ref{ggth}).

The other input for constructing warm inflation scenarios is the free
energy in the expanding environment for the model (\ref{langran}).
It already has been argued above that temperature is a good parameter
for describing the state of the radiation in the warm inflation regime.
It also follows from the above that the change in temperature can be
treated adiabatically in the thermodynamic functions, since this
requires $\Gamma_{\rm rad} \gg {\dot T}/T$, which is automatically
satisfied due to Eq. (\ref{rholth}). Therefore the free energy density
should be well represented by the Minkowski space expression,
with temperature treated as an adiabatic parameter. {}For the
model in Secs. II-IV, the free energy density is

\begin{equation}
F(\varphi_c,T)= \frac{m_T^2}{2} \varphi_c^2 + 
\frac{\lambda_T}{24} \varphi_c^4-
\frac{(N+1)\pi^2}{90} T^4 \;,
\label{freeeng}
\end{equation}

\noindent 
where the factor $N+1$, in the last term, comes 
from the functional integration over the
$\chi$ fields and the $\phi$-field's fluctuations.
Having established this to be the free energy density for the warm
inflation scenario, the other thermodynamic functions such as
pressure, energy density and entropy density can be easily obtained.

With the free energy (\ref{freeeng})
and the order parameter equation of motion, Eq.
(\ref{phiheq}), determined, the time evolution of the three unknowns:
temperature $T(t)$, scale factor $R(t)$ and order parameter
$\varphi_c(t)$, can be obtained from Eq. (\ref{phiheq}) plus any two
independent equations from Friedmann cosmology along with a self
consistency check for adiabaticity, Eq. (\ref{ddpledp}). At this point
the procedure in \cite{ab55} can be followed. However due to the
microscopic origin of this model, additional self consistency checks are
necessary for adiabaticity, given by Eq. (\ref{kinetic}) and
thermalization, Eq. (\ref{ggth}). Observationally interesting expansion
factors will require $H > {\dot \phi}/\phi$, in which case the condition
(\ref{ggth}) immediately implies the microscopic adiabatic condition
(\ref{kinetic}).

\subsection{Results}

Up to this point, the formulation of warm inflation in conjunction with
a microscopic dynamics has been general. In the remainder of this
section, some demonstrative calculations of this cosmology will be
presented based on our microscopic model. An exhaustive analysis of the
parameter space will not be performed. In this first examination, the
emphasis is to understand the interplay between the microscopic and
macroscopic physics of warm inflation for generic potentials, which in
particular, have curvature scale of order the temperature scale. For
such potentials, thermal fluctuations that displace $\varphi_c(0)$
substantially from the origin are exponentially suppressed. However, it
is such fluctuations that allow enough time, during the roll down back
to the origin, for the universe to inflate sufficiently. As such, this
elementary fact, in any case, quells significant interest in comparing
the cases we will examine to observation.

It should be noted that the order parameter in this symmetry restored
warm inflation regime is configured similar to those in the chaotic
inflation scenario \cite{linde}. However, in the chaotic inflation
scenario the potentials are ultra-flat. Such potentials permit large
fluctuations of the order parameter and in fact prefer them. The
dissipative model in this paper could be studied for the case of
ultra-flat potentials, perhaps motivated by supersymmetric model building. 
This would extend the pure quantum mechanical,
new-inflation type dynamics of chaotic inflation into the intermediate
regime discussed in \cite{bf2,rudnei}. This will not be examined here.

Proceeding with our demonstrative examination of warm inflation, let the
origin of time be the beginning of the inflationlike regime (BI) and
also the beginning of our treatment. The basic picture of the particular
warm inflation scenario studied here is as follows. At $t=0$ the initial
conditions are arranged so that the field is displaced from the origin
$\langle\phi(0)\rangle = \phi_{BI}$, the temperature of the universe is
$T_{BI}$ and since the universe is at the onset of the inflation-like
regime, by definition this means the vacuum energy density equals the
radiation energy density, $\rho_v(0) = \rho_r(0)$. For $t>0$ the field
will relax back to the origin within a strongly dissipative regime and
in the process liberate vacuum energy into radiation energy.
Simultaneously, the scale factor will undergo inflation-like expansion.
During the roll-down period, the vacuum energy first dominates until at
some point it is superseded by the radiation energy. At this point the
universe smoothly exits the inflation-like regime into the radiation
dominated regime.

{}From our model in the previous sections, we will consider the case of
$N'$ $\chi$-bosons ($\chi'$) with $g_j=g \gg f_j, j = 1 \ldots N'$ and
$N-N'$ $\chi$-bosons ($\chi$) with $g_j \ll f_j=f, j= N'+1 \ldots N$.
For this model, the dissipative dynamics of $\varphi_c$, expressed
through $m_T, \lambda_T$ and $\eta_1$, is controlled by the $N'$
$\chi$-bosons. The other $N-N'$ $\chi$-bosons only serve as additional
fields in the radiation bath. For this purpose, from Eq.
(\ref{gammachi}), for $f \geq g^2$ the $\chi$ and $\chi'$ bosons will be
equally effective in thermalizing the radiation energy.

In this paper we will examine this scenario in the regime

\begin{equation}
\frac{\lambda_T \varphi_c^4}{24} \gg \frac{m_T^2 \varphi_c^2}{2}.
\label{p4ggp2}
\end{equation}

\noindent 
and in the high temperature limit $T \gg m_T, \mu_T$. Also, for ease of
presentation, we will write the expression for $\Gamma(q)$ at
${\bf q} =0$. 
Although with these simplifications the results 
will not be cosmologically interesting,
it is a good example to demonstrate the general procedure.
In this regime, the effective
equation of motion for $\varphi_c$, from Eq. (\ref{over}), is:

\begin{equation}
\frac{d \varphi_c}{dt} = -\frac{B_2}{4} \varphi_c \;,
\end{equation}

\noindent
where

\begin{equation}
B_2 \equiv \frac{2 \lambda_T}{3 \eta_1} \approx
\frac{ \pi T_{BI}\lambda_T}{72 N' {\rm ln} {(2T_{BI}/\mu_{T_{BI}})}}.
\end{equation}

\noindent
{}Formally the Friedmann cosmology for the
warm inflation scenario associated with the above
equation was called the quadratic limit in \cite{ab55}.

The macroscopic and microscopic requirements of warm inflation will
imply various parametric constraints which are as follows. Eq.
(\ref{p4ggp2}) will be satisfied by requiring:

\begin{equation}
\frac{\lambda_T \varphi_{BI}^4}{24} =
{\rm r} \frac{m_T^2 \varphi_{BI}^2}{2}\;,
\label{p4erp2}
\end{equation}

\noindent
where the parameter ${\rm r} \gg 1$ has been introduced. As shown in
\cite{ab55}, throughout the inflation-like period until just before it
ends, the temperature drops slightly faster than $\phi$. As such, the
thermal mass term, $m_T^2 \phi^2/2 \sim T^2 \phi^2$, will continue to
satisfy Eq. (\ref{p4ggp2}) given that initially it does.

The number of e-folds, $N_e$ obtained during the
roll-down is, from \cite{ab55},

\begin{equation}
N_e \approx \frac{2H}{B_2}\;,
\label{rne}
\end{equation}

\noindent
where

\begin{equation}
H = \sqrt{\frac{8 \pi \lambda_T \varphi^4_{BI}}{72 m_p^2}} \;,
\end{equation}

\noindent
with $m_p$ the Planck mass.
The microscopic condition, Eq. (\ref{ggth}), requires

\begin{equation}
\frac{g^4 T}{192 \pi} \gg
\frac{\sqrt{\pi \lambda_T} \varphi_{BI}^2}{3 m_p}.
\label{micro}
\end{equation}

\noindent
The threshold condition for inflation, $\rho_v(0)=\rho_r(0)$, implies

\begin{equation}
\frac{(N+1) \pi^2}{30}T^4_{BI} = \frac{\lambda_T \varphi^4_{BI}}{24}.
\label{rverr}
\end{equation}

\noindent
Finally, the validity of the perturbative derivation in the previous
sections will require

\begin{equation}
g, \lambda, \frac{m_T}{T},
\lambda_T  < 1 \;,
\label{parml1}
\end{equation}

\noindent
where $m_T^2 \sim (\lambda + 2 N' g^2)T^2/24$ and 
$\lambda_T \approx \lambda - \frac{3 \sqrt{3}}{2 \pi} N' g^3$.

Eq. (\ref{micro}) can be turned into an equality, in which case, along
with Eqs. (\ref{p4erp2}), (\ref{rne}), and (\ref{rverr}), they determine
the boundary of the allowed parameter space. Thus there are six
constraining equations for the eleven quantities
$\lambda,g,\lambda_T,m_T,m_p,\varphi_{BI},N,N',N_e,{\rm r},$ and
$T_{BI}$. We will let $T_{BI}$ set the overall scale and will fix
$N',N_e,{\rm r},g$. Then, based on the constraint equations, this
determines the remaining parameters. In particular, we have for
$\lambda_T$: 

\begin{equation}
\lambda_T \approx \frac{3 N' g^4 \; {\rm ln} (2 \sqrt{12}/g)}{4 N_e \pi^2}.
\label{rlamt}
\end{equation}

\noindent
This expression is suggestive of the case 3 analyzed in Sec. IV.
{}For such, taking then 
$m^2_T \sim \left(3g \sqrt{3}/2 \pi +2 \right) N' g^2 T^2/24$,
we get the additional parameters:

\begin{equation}
\varphi_{BI} \approx \frac{2 \pi T_{BI}}{g} 
\sqrt{\frac{\left(3 g \sqrt{3}/2 \pi + 2 \right) N_e 
{\rm r}}{6 \;{\rm ln} (2 \sqrt{12}/g)}} \;,
\label{rphibi}
\end{equation}

\begin{equation}
N+1 \approx \frac{5 {\rm r}^2 N' N_e}{12 \;{\rm ln} (2 \sqrt{12}/g)}
\left(\frac{3\sqrt{3}}{2 \pi} g + 2 \right)^2 \; ,
\label{rnnp}
\end{equation}

\noindent
and

\begin{equation}
m_p \approx \frac{192 \pi^2 {\rm r} T_{BI}}{9g^4}
\left(\frac{3\sqrt{3}}{2 \pi} g + 2 \right)
\sqrt{\frac{3 \pi N' N_e}{ {\rm ln} (2 \sqrt{12}/g)}} \;.
\label{rmpl}
\end{equation}

Based on these equations, it is not difficult to find parametric regimes
in which the warm inflation scenario is realized, but it is only for
$N_e< 1$.  As such, this simple case has no observational relevance.
There are a few improvements that could be made to our analysis that
would increase $N_e$.
Firstly our estimates above ignore the
effects of the thermal mass term, 
$m_T^2 \varphi_c^2/2$, on the dynamics.  Its contribution to the
energy density and pressure are $\rho_{m_T}= -m_T^2 \varphi_c^2/2 $
and $p_{m_T} = \rho_{m_T}$ respectively.  Thus it helps the $\varphi_c^4$
term to drive inflation.  Secondly, recall that the dissipative
coefficient $\eta_1 \sim 1/T$.  In the above analysis, we fixed 
$T=T_{BI}$. However, during the roll-down, temperature does fall by a
factor of order ten, which in turn would increase $\eta_1$.
Finally the parameter regime could be extended to include both positive
and negative $\phi-\chi$ couplings, such as the ASR case described in
subsection IV.A.  As noted there, in this regime the duration of
overdamped motion can be increased significantly within the perturbative
amplitude expansion. This directly corresponds to increasing the
e-folds $N_e$.

A more elementary modification is to extend the region of validity to
larger displacements of $\varphi_c$. The extension to
this larger regime can be treated by a summation of the complete
one-loop series as outlined in subsection IV.B.
Overdamped motion for much larger displacements of $\varphi_c$ can also
be attained by a modification to the $\phi-\chi$ interaction in
Eq. (\ref{Nfields}) as
\begin{equation}
\sum_{j=1}^N \frac{g^2}{2} \phi^2 \chi_j^2 \rightarrow
\sum_{j=1}^N \frac{g^2}{2} (\phi - M_j)^2 \chi_j^2 .
\end{equation}
In this distributed mass model (DMM), a given $\chi_j$-field is thermally
excited when its effective mass $g^2(\varphi_c-M_j)^2 < T^2$. The
contribution from the thermally excited $\chi$-fields to the effective
dynamics of $\varphi_c$ is similar to our calculations in sections
II-IV.  As such, an effective equation of motion similar to Eq. (4.4) can
be obtained for this modified model.  Given an appropriate distribution
of mass coefficients $M_j$ along the path of $\varphi_c$, from an
arbitrarily large initial displacement $\varphi_{BI}$, $\varphi_c(t)$
could undergo overdamped motion along its entire path.  Details will be
presented elsewhere on the warm inflation scenario which considers these
various cases.

{}For this ``symmetry restored'' case, initial fluctuations of
$\varphi_c(t)$, $\varphi_{BI}$, are strongly suppressed with probability
$\exp(-{\rm volume} \ \rho/T_{BI}) \sim \exp[- (128 \pi)^3 N N_e {\rm
r}^2/ g^{12}]$, where the volume $ \sim 1/H^3$. {}The most optimistic
initial conditions have probability $ \sim \exp[-1 \times 10^{7}]$.
Thus, unless a viable mechanism is found to justify a large enough
initial value of $\varphi_c$, the regime investigated here may not be
very relevant for practical applications of warm inflation. In any case,
the microscopic dynamics of the symmetry restored regime investigated
here is similar to more realistic scenarios in the symmetry broken
regime, where the field has an average initial value close or identical
to zero. An important difference being that the initial state in the
latter case has no Boltzmann suppression.

This section has made an initial examination of treating strong
dissipation from first principles during a de Sitter expansion regime.
Further results will be presented elsewhere as well as a calculation
similar to this one, for the symmetry broken case.

\section{Conclusion}
\label{sec6}

In this paper a microscopic quantum field theory model has been
presented, describing overdamped motion of a scalar field. Commonly,
such behavior is treated phenomenologically by Ginzburg-Landau order
parameter kinetics. Our model provides a first principle explanation of
how kinetics equivalent to the Ginzburg-Landau type, which is first
order in time, arise for inherently second order dynamical systems. The
microscopic treatment of this problem, in principle, should be well
controlled, due to its fundamental reliance on the adiabatic limit, and
our model exemplifies this expectation.

The calculational method for treating dissipation in this paper has one
distinct difference from several other related works. In our
calculation, we consider the effect of particle lifetimes in the
effective equation of motion. To our knowledge, this effect has been
discussed in only a few works in the past \cite{GR,hs1,morikawa}. 

Secs. II-IV presented a general, flat-space treatment, which offers a
microscopic justification to the often used limit of diffusive
Ginzburg-Landau scalar field dynamics. We have shown how it is possible
to obtain an effective evolution for the scalar field which is
first-order in time, due to its own thermal dissipation effects,
interpreted microscopically as its decay into many quanta. In a sense,
the field acts as its own brakes, the slowing of its dynamics being
attributed to the highly viscous medium where it propagates, a densely
populated sea of its own decay products.

The application that we considered in Sec. V was in expanding spacetime,
for the cosmological warm inflation scenario. Although we did not
formally derive the extension of our flat-space model of Secs. II-IV to
an expanding background, we did present heuristic arguments that
validate this extension for the special needs of warm inflation. The
results of the simple analysis in Sec. V are strongly dependent on
initial conditions and may be difficult to implement for models of
observational interest. Nevertheless, these results will provide useful
guidance both for modifications of this model and for our next study of
the symmetry broken regime.

The direct significance of the present study to inflationary cosmology
would be to the initial state problem \cite{muw} in scenarios during
symmetry breaking. The initial conditions required for warm inflation in
the symmetry broken case are similar to new inflation. The requirement
is a thermalized inflaton field, which at the onset of the warm
inflation regime is homogeneous with expectation value $\langle \phi
\rangle_{\beta}=0$. Although we have made no detailed application of our
results to this problem, some general features are evident from the
analysis in Sec. V. In particular, both the suppression of large
fluctuations and thermalization are mutually consistent with strong
dissipative dynamics. Many of the difficulties that have been discussed
\cite{muw,cl} in association with the initial state problem, are
eliminated in the strong dissipative regime. In addition, the damping of
fluctuations should simplify the formal problem of coupling this model
to classical gravity. Thus, the strong dissipative regime appears to
have the correct features both to carry the universe into an
inflation-like phase and then to smoothly exit into a hot big-bang
regime.

\acknowledgements

We thank R. Holman for helpful discussions. AB was supported by a
Department of Energy grant. MG was partially supported at Dartmouth
College by the National Science Foundation through a Presidential
Faculty Fellows Award no. PHY-9453431 and by the National Aeronautics
and Space Administration grant no. NAGW-4270.
MG thanks both Fermilab and the Osservatorio di Roma for
their kind hospitality during the completion of this work. ROR was
partially supported by Funda\c{c}\~ao de Amparo \`a Pesquisa no
Estado do Rio de Janeiro - FAPERJ and by Conselho Nacional de 
Desenvolvimento Cient\'{\i}fico e Tecnol\'ogico - CNPq.

\appendix

\section{}

We now give a brief overview of the calculation of the imaginary part of
the two-loop, setting sun, self-energy terms in (\ref{selfphi}) and 
(\ref{selfchi}). Let us express generically those diagrams in terms of
field propagators with masses $m_s$ and $m_t$ and external lines
of type $s$. {}For an interaction between $s$ and $t$ fields of the
form $g^2_{s,t}/2\; s^2 t^2$ (for $s=t=\phi \; (\chi_j)$, 
$g^2_{s,t}=\lambda/12 \; (f_j/12)$ and for $s = \phi,\, t=\chi_j$,
$g^2_{s,t} = g_j^2$) the imaginary part of the two-loop sunset 
diagram for the $s$ field can be expressed by (see for example,
in Ref. \cite{jeon2}, Appendix G and also Wang and Heinz in 
\cite{heinz}):

\begin{eqnarray}
\begin{picture}(320,28) 
\put(-70,0){${\rm Im}\Sigma_s (q)\; = $}

\thicklines
\put(-5,0){${\rm Im} \;\left[ \hspace{3cm} \right]\;=$}
\put(30,0){\line(1,0){60}}
\put(60,0){\circle{30}}
\put(57,20){$t$}
\put(57,3){$s$}
\put(57,-25){$t$}
\end{picture} 
\nonumber
\end{eqnarray}

\vspace{0.25cm}

\begin{eqnarray}
\lefteqn{= S_{s,t} \: g_{s,t}^4 \left(1 - e^{-\beta E_q^s}\right)
\sum_{\sigma=\pm 1} \sigma_{k+q} \sigma_{l} \sigma_{k+l}
\frac{1}{8 (2 \pi)^5} 
\int \frac{d^3 k d^3 l}{E_l^t E_{k+q}^s E_{k+l}^t}
\left[ 1 + n(\sigma_{k+q} E_{k+q}^s)\right] }
\nonumber \\
& & \times \left[ 1 + n(\sigma_{l} E_{l}^t)\right]
\left[ 1 + n(\sigma_{k+l} E_{k+l}^t)\right]
\delta(E_q^s - \sigma_l E_l^t - \sigma_{k+q} E_{k+q}^s - 
\sigma_{k+l} E_{k+l}^t )\:,
\hspace{2.5cm}  
\label{B1}
\end{eqnarray}

\noindent
where $E_k^s = \sqrt{{\bf k}^2 + m_s^2}$, $n(E)$ is the bose-distribution
function and 
$S_{s,t}$ is a symmetry factor: for $s=t$, $S=12$ and for
$s \neq t$, $S=1$. By expanding in the sum in $\sigma$ and retaining 
only the on-shell, energy conserving processes (corresponding to
the scattering processes $st \to st$, $ts \to
ts$), we obtain the result:

\begin{eqnarray}
{\rm Im} \Sigma_s (q) = & S_{s,t} & g^4_{s,t}
\frac{\left(1+\frac{1}{2}\delta_{s,t}\right)}{4
(2 \pi)^5} \left(1 - e^{-\beta E_q^s}\right)
\int \frac{d^3 k d^3 l}{E_l^t E_{k+q}^s E_{k+l}^t}
\left[ 1 + n(E_{k+q}^s)\right]
\left[ 1 + n(E_{l}^t)\right]
n(E_{k+l}^t) \nonumber \\
& \times &  \delta(E_q^s - E_l^t - E_{k+q}^s + E_{k+l}^t )\:,
\label{B2}
\end{eqnarray}

\noindent
with $\delta_{s,t} =1$
for $m_s=m_t$ and $\delta_{s,t} =0$ for $m_s \neq m_t$. ${\rm Im}
\Sigma (q)$ for $s=t$ has been obtained in details in Refs. \cite{jeon2}
and \cite{heinz}. In particular, Wand and Heinz in \cite{heinz} have
discussed and obtained in detail the kinematic limits, for the 
$\lambda \phi^4$ model, of the integration
on the momenta in (\ref{B2}), implicit in the Dirac delta-function.
Here we obtain the results for the case of $m_s \neq m_t$. 
In (\ref{B2}), by defining in the three-dimensional momentum
integrations the angular differentials $d (\cos \theta_k) =
d E^s_{k+q} \: E^s_{k+q}/
(k q) $ and  $d (\cos \theta_l) = d E^t_{k+l} \:E^t_{k+l}/
(k l)\; $ 
($k,q,l = |{\bf k}|,|{\bf q}|,|{\bf l}|)$, we are then able to
perform the angular integrals in 
(\ref{B2}).
{}From the constraint in the integration limits
for $k$ and $l$, which comes from the delta-function, we obtain the
following result for ${\rm Im} \Sigma_s (q)$,

\begin{eqnarray}
{\rm Im} \Sigma_s (q)  = & S_{s,t} & g^4_{s,t} 
\frac{T \left(1+\frac{1}{2}\delta_{s,t}\right)}{4
(2 \pi)^3 q} \left(1 - e^{-\beta E_q^s}\right) \left\{
\int_0^\infty d k 
\int_{-u(k,q)}^{u(-k,q)}
\frac{l dl}{E_l^t} \left[ 1 + n(E_q^s - E_l^t)\right]
\left[ 1 + n(E_{l}^t)\right] \right. \nonumber \\
& \times & \left. {\rm ln} \left( \frac{e^{\beta E_{k+l}^t} -
1}{e^{\beta E_{k+l}^t}- e^{\beta (E_{l}^t- E_q^s)}} \right)
+ \left( \int_{-q}^\infty dk \int_{u(k+q,q)}^\infty 
\frac{l dl}{E_l^t} - \int_{q}^\infty dk \int_{u(-k+q,q)}^\infty 
\frac{l dl}{E_l^t} \right) \right. \nonumber \\
& \times & \left. \left[ 1 + n(E_q^s - E_l^t)\right]
\left[ 1 + n(E_{l}^t)\right] {\rm ln} \left( \frac{e^{\beta E_{k}^s} -
1}{e^{\beta E_{k}^s}- e^{\beta (E_{q}^s- E_l^t)}} \right)
\right\} \;,
\label{B3}
\end{eqnarray}

\noindent
where, in the above expressions, the function $u(k,q)$ is given by

\begin{eqnarray}
u(k,q) = & \frac{1}{2} & \left\{ k - \frac{1}{k m_s} \left[ \sqrt{(k-q)^2 +
m_s^2} - \sqrt{q^2+m_s^2} \right] \left(k^2 m_s^2 + 2 m_s^2 m_t^2 + 2
q^2 m_t^2 \right. \right. \nonumber \\
& - & \left. \left.
2 kq m_t^2 + 2 m_t^2 \sqrt{(k-q)^2 + m_s^2} \sqrt{q^2+m_s^2}
\right)^{\frac{1}{2}} \right\}\;.
\label{B4}
\end{eqnarray}

\noindent
In the first term of (\ref{B3}), we can make the change of integration
variables $k+l= k'$, $l= l'$, to obtain
$l'dl'/E_{l'}^t = dE_{l'}^t$. Doing the same for
the remaining integrals, $ldl/E_{l}^t = dE_{l}^t$, relabeling
$l'$ and $k'$ back to $l$, $k$, and performing a new change of integration
variables, $y = e^{-\beta E_l^t}$, $dy = -\beta e^{-\beta E_l^t}
dE_l^t$, we are able to compute the $y$ integrations. {}From this
point on, the integrations are equivalent to the ones
in \cite{jeon2}, once we change the integration limits in the $k$ and $l$
integrals, and take into account the function $u(k,q)$, Eq. (\ref{B4}).
The results shown in Sec. II for $\Gamma_\phi$,
$\Gamma_{\chi_j}$, Eqs. (\ref{gammaphi}) and (\ref{gammachi}), are
obtained once the limit of vanishing $q$ ($|{\bf q}| =0$), is taken in
the above equations.

\section{}

Let us now compute the expression on the LHS of (\ref{dissterm}) in the 
adiabatic approximation, stressing the need for regularizing the propagators
with quasi-particle lifetimes. Let us write the LHS of
(\ref{dissterm}) in terms of the adiabatic series:

\begin{eqnarray}
\lefteqn{\int d^4 x' \varphi_c^2 (x') \rm{Im} 
\left[G^{++}\right]_{x,x'}^2 \theta (t-t')} \nonumber \\
& & = \sum_n \int d^3 x' \int \frac{d^3 k}{(2 \pi)^3} e^{i {\bf k}.({\bf
x} - {\bf x}')} \frac{1}{n!} \left. \frac{\partial^n \left(\varphi_c^2
\right)}{\partial t^{\prime \, n}} \right|_{t'=t} \int \frac{d^3 q}{(2
\pi)^3} \int_{-\infty}^t d t' (t'-t)^n \nonumber \\
& & \times {\rm Im} \left[ G^{++} ({\bf q}, |t-
t'|) G^{++} ({\bf q} - {\bf k}, |t-t'|) \right] \;,
\label{A1}
\end{eqnarray}

\noindent 
where $G^{++} ({\bf q}, t-t')$ can be read from (\ref{G of
k}) and (\ref{G><}) and it refers generically to the $\phi$ or $\chi_j$
field propagators. The approximation of considering
a homogeneous field, $\varphi_c
\equiv \varphi_c (t)$, is equivalent to taking the limit ${\bf k} \to 0$
for the external momentum in $G^{++} ({\bf q}- {\bf k}, |t-t'|)$.
In this case the
$x'$ and $k$ integrations in (\ref{A1}) can be done trivially. However, 
it is known that
taking the limit of zero
external momentum ($k_\mu \to 0$) \cite{gross,deriv} in self-energy
expressions, which are given in terms of products of non-local propagators, 
can be problematic. This
is related to the non-analyticity of these expressions around the
origin. In particular, Gross, Pisarski, and Yaffe in \cite{gross}
argue that a correct way of taking the limit $k_\mu \to 0$, at finite
temperatures, is to first take $k_0 \to 0$ and then ${\bf k} \to 0$.
They also argue that the non-analyticity problem, which comes from a failure
to do a self- consistent calculation, would be eliminated once
fully-dressed propagators are taken consistently into account. We note, in
particular, that for fully-dressed propagators the decay width $\Gamma$ 
works as a
regulator.

In (\ref{A1}), the limit $k_0 \to 0$ is implicit in the adiabatic
approximation, where the fields are required to change slowly in time,
while the limit ${\bf k} \to 0$ is implemented by approximating the fields
to be homogeneous. In evaluating (\ref{A1}), we will first
compute the time integral, expand in terms
of the ``regulator'' $\Gamma$, and then 
finally take the limit ${\bf k} \to 0$. This is the opposite of what was done
in obtaining Eq. (\ref{dissterm}).  {}From this lesson we will see both
the importance of considering fully-dressed propagators and how consistent
results can be obtained once the limits are taken in the correct order. 
This will be crucial when
evaluating the dissipation contribution ($n=1$, in (\ref{A1})). 

Consider the time integral on the RHS of Eq. (\ref{A1}) and expand it to
first order in the ``regulator'', $\Gamma$
(which
is of order $\lambda^2$, $g^4$). Higher order terms in $\Gamma$ need to
be considered in conjunction with higher order loop terms, for
consistency. {}For $n=0$, the zeroth-order in the adiabatic
approximation, we obtain

\begin{eqnarray}
\lefteqn{\int_{- \infty}^t d t' {\rm Im} \left[G^{++} ({\bf q}, |t-
t'|) G^{++} ({\bf q} - {\bf k}, |t-t'|) \right]} \nonumber \\
& & \simeq \frac{1}{4 \omega_{{\bf q}} \omega_{{\bf q} - {\bf k}}}
\left[ - \frac{1}{\omega_{{\bf q}} + \omega_{{\bf q} - {\bf k}}} +
2  \frac{\omega_{{\bf q} - {\bf k}} n(\omega_{{\bf q}}) -  \omega_{{\bf q}} 
n(\omega_{{\bf q} - {\bf k}})}{\omega_{{\bf q}}^2 -
\omega_{{\bf q} - {\bf k}}^2} \right] + {\cal O} \left[ \left(
\frac{\Gamma_{{\bf q}} + \Gamma_{{\bf q} - {\bf k}}}{\omega_{{\bf q}} +
\omega_{{\bf q} - {\bf k}}} \right)^2 \right] \;.
\label{A2}
\end{eqnarray}

\noindent
Taking the limit ${\bf k} \to 0$, we obtain

\begin{equation}
\lim_{{\bf k} \to 0} \int_{- \infty}^t d t' {\rm Im} \left[G^{++} 
({\bf q}, |t-t'|) G^{++} ({\bf q} - {\bf k}, |t-t'|) \right] 
\to - \frac{\left[1 + 2
n(\omega_{{\bf q}}) \right]}{8 \omega_{{\bf q}}^3} - \beta
\frac{n(\omega_{{\bf q}}) \left[1 + n (\omega_{{\bf q}}) \right]}{4
\omega_{{\bf q}}^2} + {\cal O} \left( \frac{\Gamma^2}{\omega^2} \right)
\;.
\label{A3}
\end{equation}

\noindent
As expected,
(\ref{A3}) is recognized to be the usual 1-loop correction to the
quartic coupling constant.

{}For $n=1$, the first order in the adiabatic approximation, 
which will give the dissipation coefficient, 
we obtain (again retaining terms up to
first order in $\Gamma$)

\begin{eqnarray}
\lefteqn{\int_{- \infty}^t d t' (t'-t) {\rm Im} \left[G^{++} ({\bf q}, |t-
t'|) G^{++} ({\bf q} - {\bf k}, |t-t'|) \right] } \nonumber \\
& & \simeq - \left[ 1 + 2 n(\omega_{{\bf q}})\right] \frac{ 3
\omega_{{\bf q}}^2 + \omega_{{\bf q}-{\bf k}}^2}{\omega_{{\bf q}} 
\left(\omega_{{\bf q}}^2 - \omega_{{\bf q}-{\bf k}}^2 \right)^2
\left(\omega_{{\bf q}} - \omega_{{\bf q}-{\bf k}} \right)}
\frac{\Gamma_{{\bf q}} + \Gamma_{{\bf q} - {\bf k}}}{\omega_{{\bf q}} +
\omega_{{\bf q} - {\bf k}}} + ( \omega_{{\bf q}} \rightleftharpoons
\omega_{{\bf q}-{\bf k}}) \nonumber \\
& & - \beta \Gamma_{{\bf q}} n(\omega_{{\bf q}}) 
\left[1 +n(\omega_{{\bf q}})\right] \frac{1}{\left(\omega_{{\bf q}}^2 - 
\omega_{{\bf q}-{\bf k}}^2 \right)^2} + 
( \omega_{{\bf q}} \rightleftharpoons
\omega_{{\bf q}-{\bf k}}) + {\cal O} \left[ \left(
\frac{\Gamma_{{\bf q}} + \Gamma_{{\bf q} - {\bf k}}}{\omega_{{\bf q}} +
\omega_{{\bf q} - {\bf k}}} \right)^3 \right] \;.
\label{A4}
\end{eqnarray}

\noindent
We can see, in contrast with (\ref{A2}), that the
limit ${\bf k} \to 0$ is divergent.
This divergence is reminiscent of the on-shell singularity 
which is present
in the
integral in (\ref{A4}) when bare propagators are used, thus showing the
importance of $\Gamma$ as a regulator. By first taking the homogeneous
limit ${\bf k} \to 0$ and then expanding in $\Gamma$, we obtain the
result given in the text, which is the first term 
on the RHS in (\ref{dissterm}).

The same calculation can be performed for the
$n=2$ case, the second order in the adiabatic expansion, from which we
obtain

\begin{equation}
\lim_{{\bf k} \to 0} \int_{- \infty}^t d t' (t'-t)^2
{\rm Im} \left[G^{++} ({\bf q}, |t-
t'|) G^{++} ({\bf q} - {\bf k}, |t-t'|) \right] \to \frac{1 + 2
n(\omega_{{\bf q}})}{16 \omega_{{\bf q}}^5} + {\cal O} \left(
\frac{\Gamma^2}{\omega^2} \right) \;.
\label{A5}
\end{equation}

\noindent
This result is consistent with a recent calculation in Ref.
\cite{deriv}, which addresses the time derivative expansion of the effective
action, for a given scalar field model. The authors of \cite{deriv}
also discuss their work in the context of the non-analyticity problem in
finite temperature QFT. In fact, we note that the second order term
in the adiabatic approximation can be associated with the first order
term in the time derivative expansion of the 
effective action, $\sim {\cal Z}(\varphi) (\partial_t \varphi)^2$
(for a time-dependent, space-homogeneous 
field configuration).

\newpage

\begin{center}
FIGURE CAPTIONS
\end{center}

\vspace{0.5cm}

{\bf Figure 1}: ${\rm Im} \Sigma_{\chi} 
({\bf q}, \omega_{\chi} ({\bf q}))$ normalized by its $|{\bf q}| =0$ 
value, for different values of masses and space momentum.

{\bf Figure 2}: The dissipation coefficient $\eta_1$ computed 
(for $N=25$) with ${\rm Im} \Sigma_{\chi} 
({\bf q}, \omega_{\chi} ({\bf q}))$ and with the approximation
$|{\bf q}| =0$ for the imaginary part of the self-energy.

{\bf Figure 3}: Results for the adiabatic condition, Eq.
(\ref{kinetic2}). The dashed lines correspond to
$| \dot{\varphi}_c/\varphi_c|$, for different values for
$\varphi_c$. The solid line corresponds to $\Gamma_{\chi} (q)$, evaluated
at $|{\bf q}| = T$. All cases shown are for $N=25$. 
The region satisfying the
adiabatic condition is the intersection of the region above the dashed
lines with the region below the solid line.
Fig. 3a is for $\lambda=g^2$, Fig. 3b is for $\lambda=g$
and Fig. 3c is for $\lambda_T=g^4$.

\begin{figure}[b]
\epsfysize=18cm 
{\centerline{\epsfbox{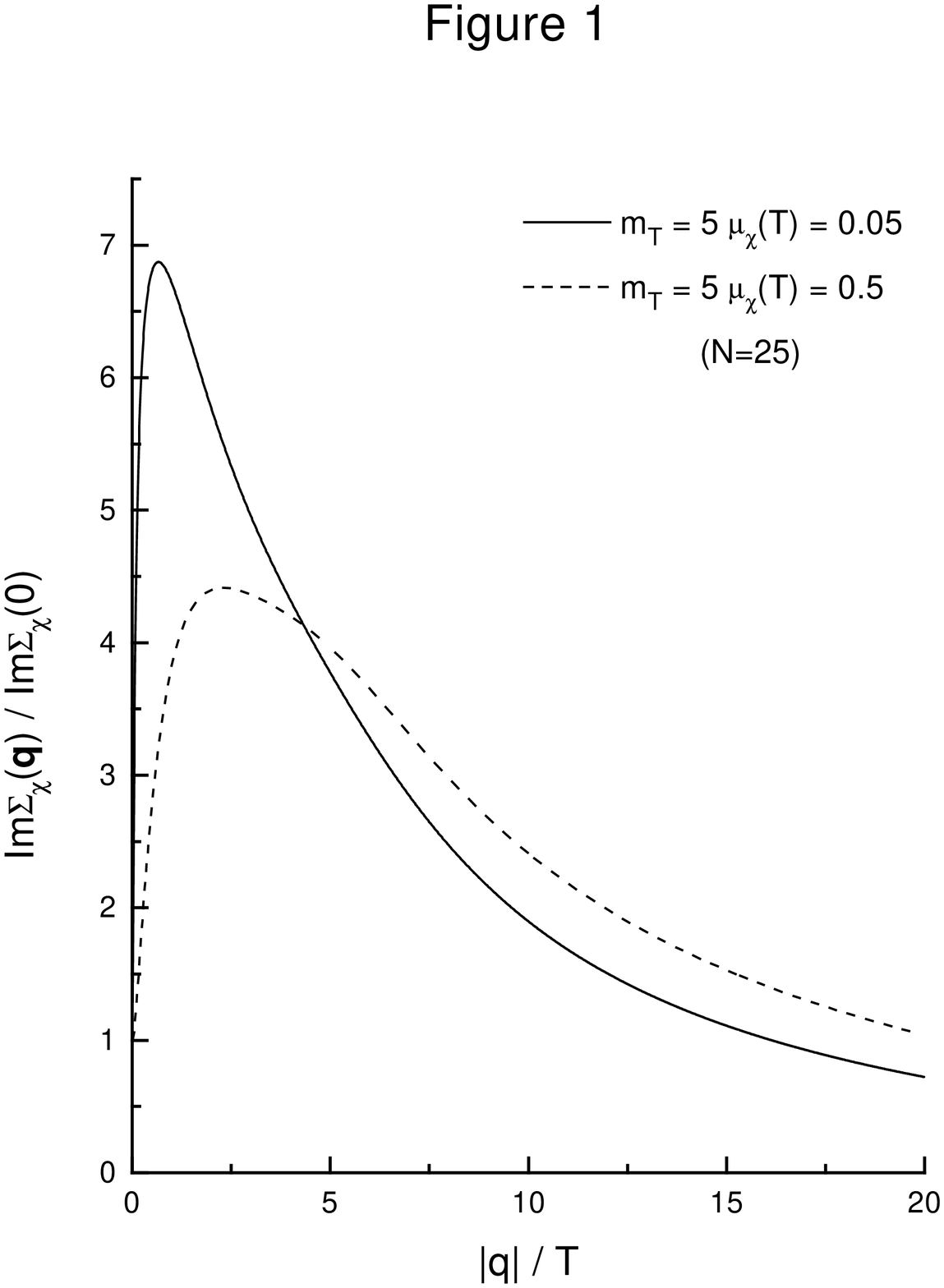}}}

\vspace{1cm}

\end{figure}

\newpage

\begin{figure}[b]
\epsfysize=18cm 
{\centerline{\epsfbox{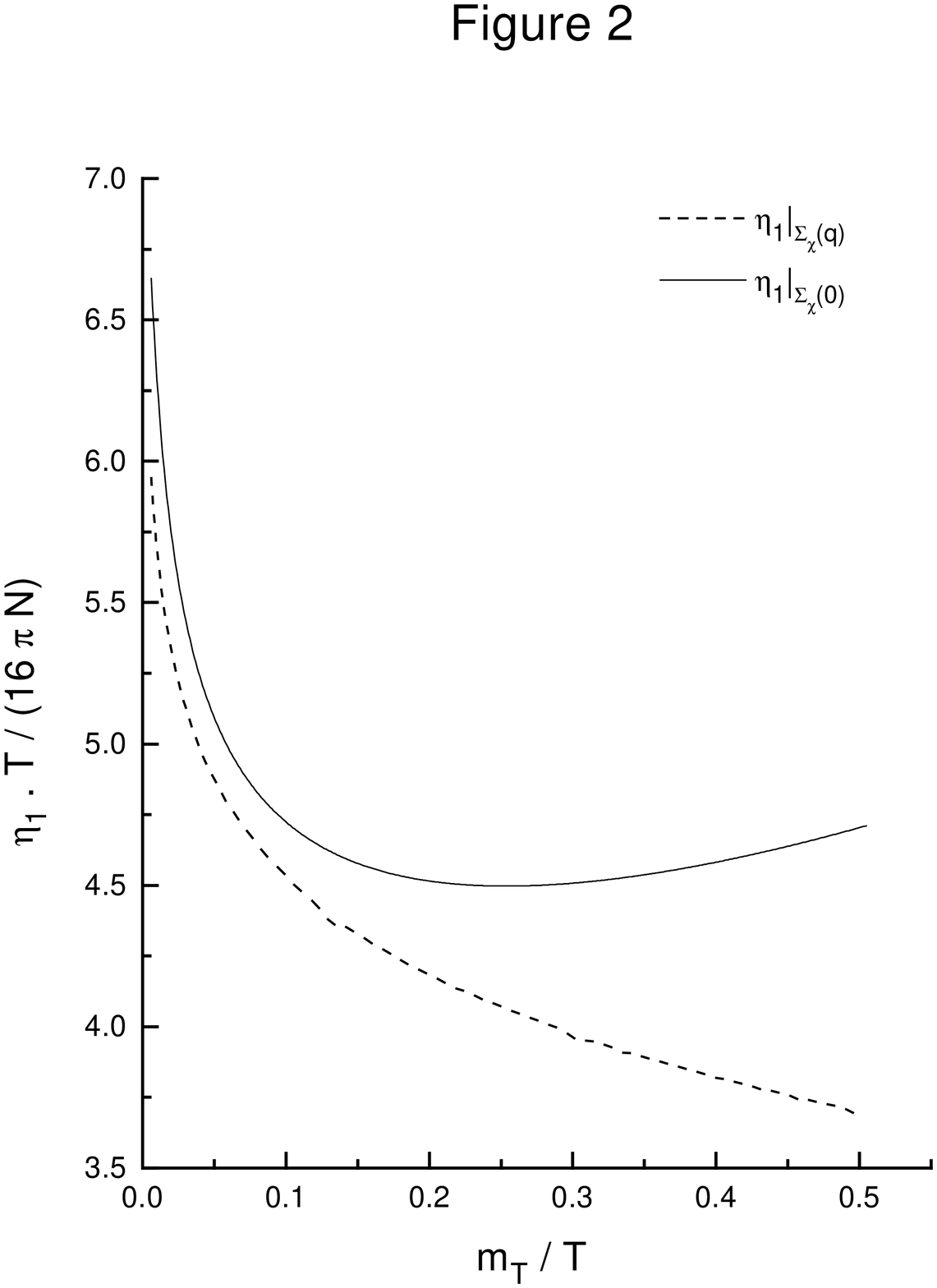}}}

\end{figure}

\newpage

\begin{figure}[b]
\epsfysize=18cm 
{\centerline{\epsfbox{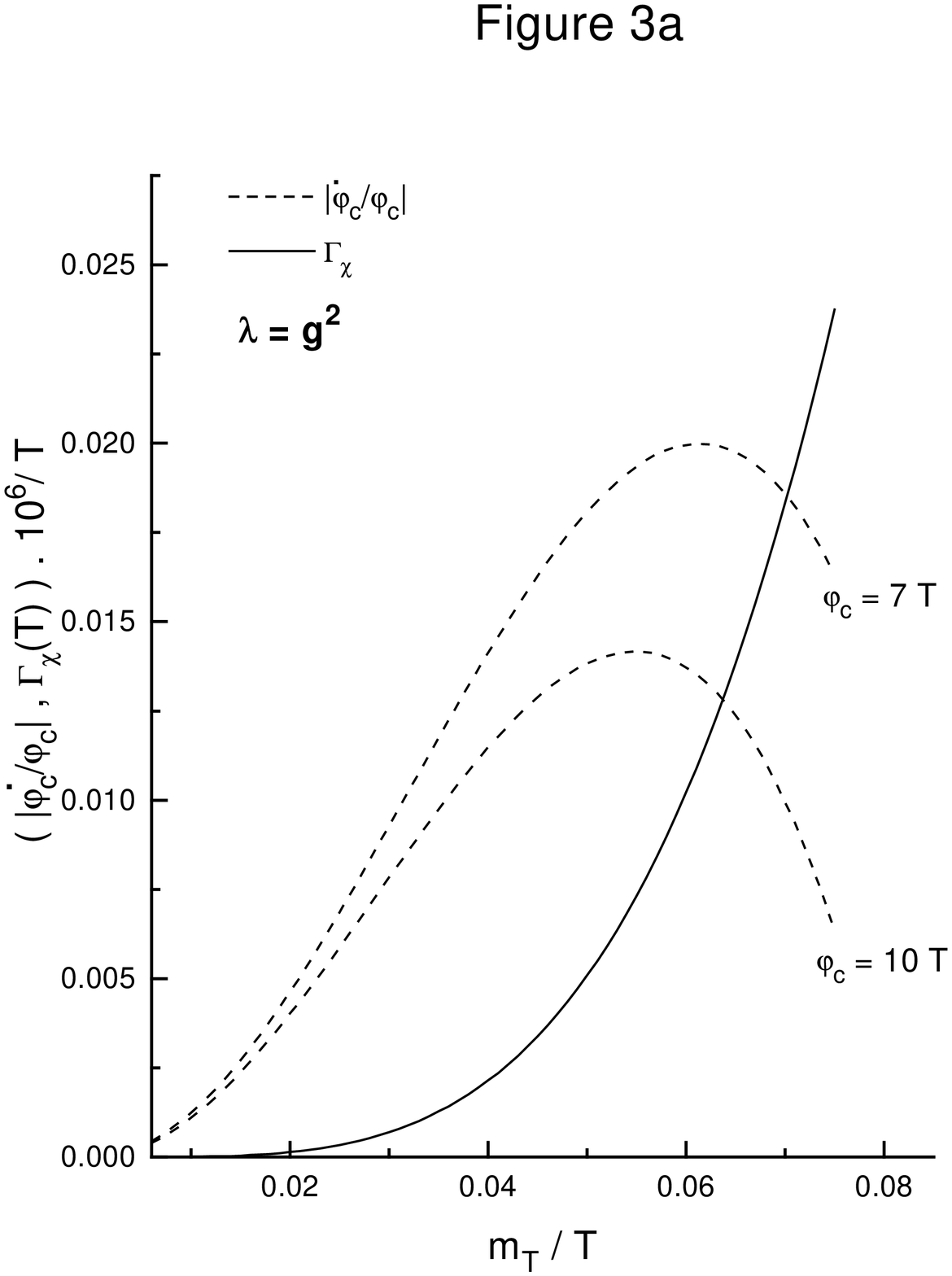}}}
\end{figure}

\newpage

\begin{figure}[b]
\epsfysize=18cm 
{\centerline{\epsfbox{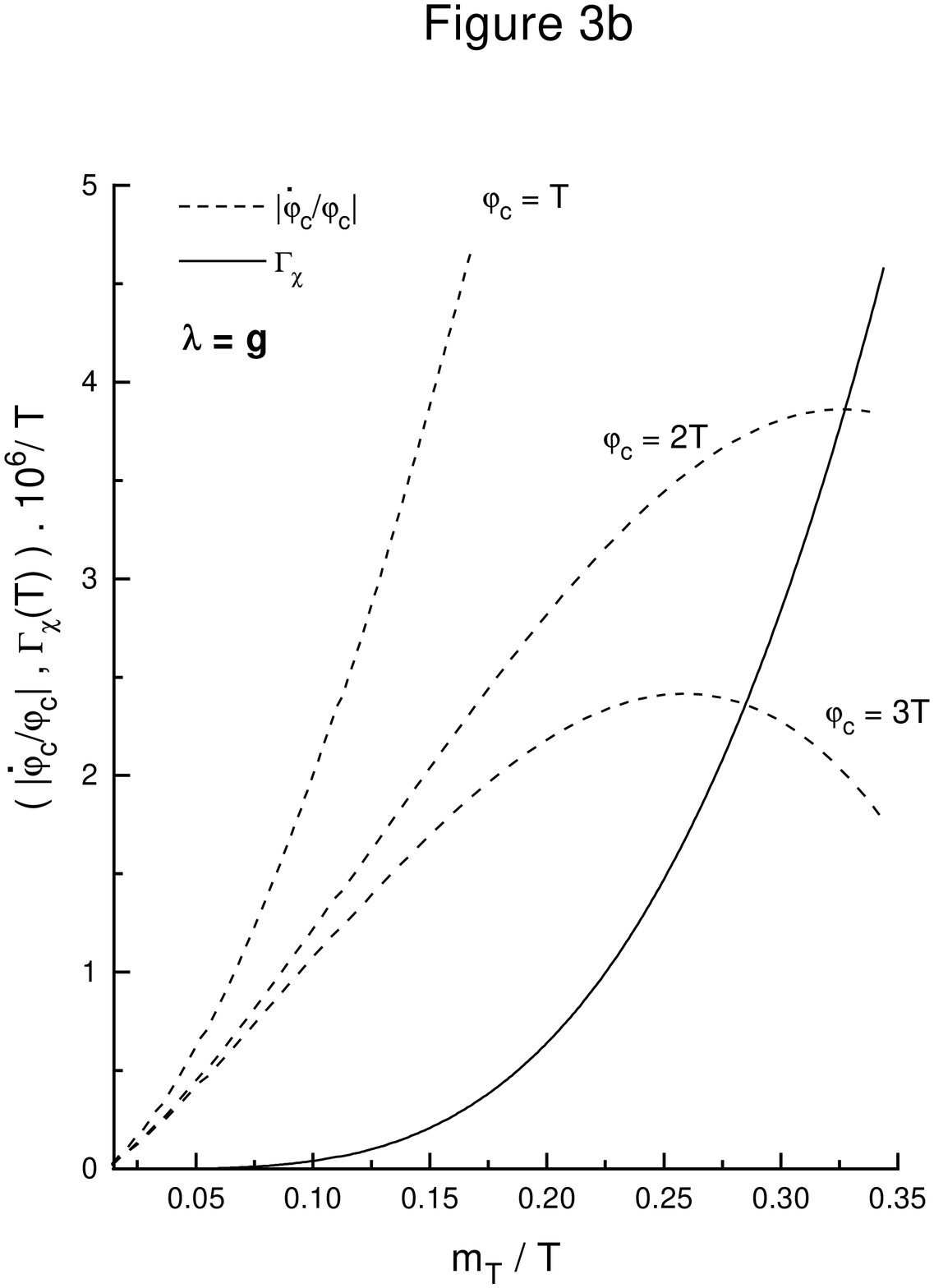}}}
\end{figure}

\newpage

\begin{figure}[b]
\epsfysize=18cm 
{\centerline{\epsfbox{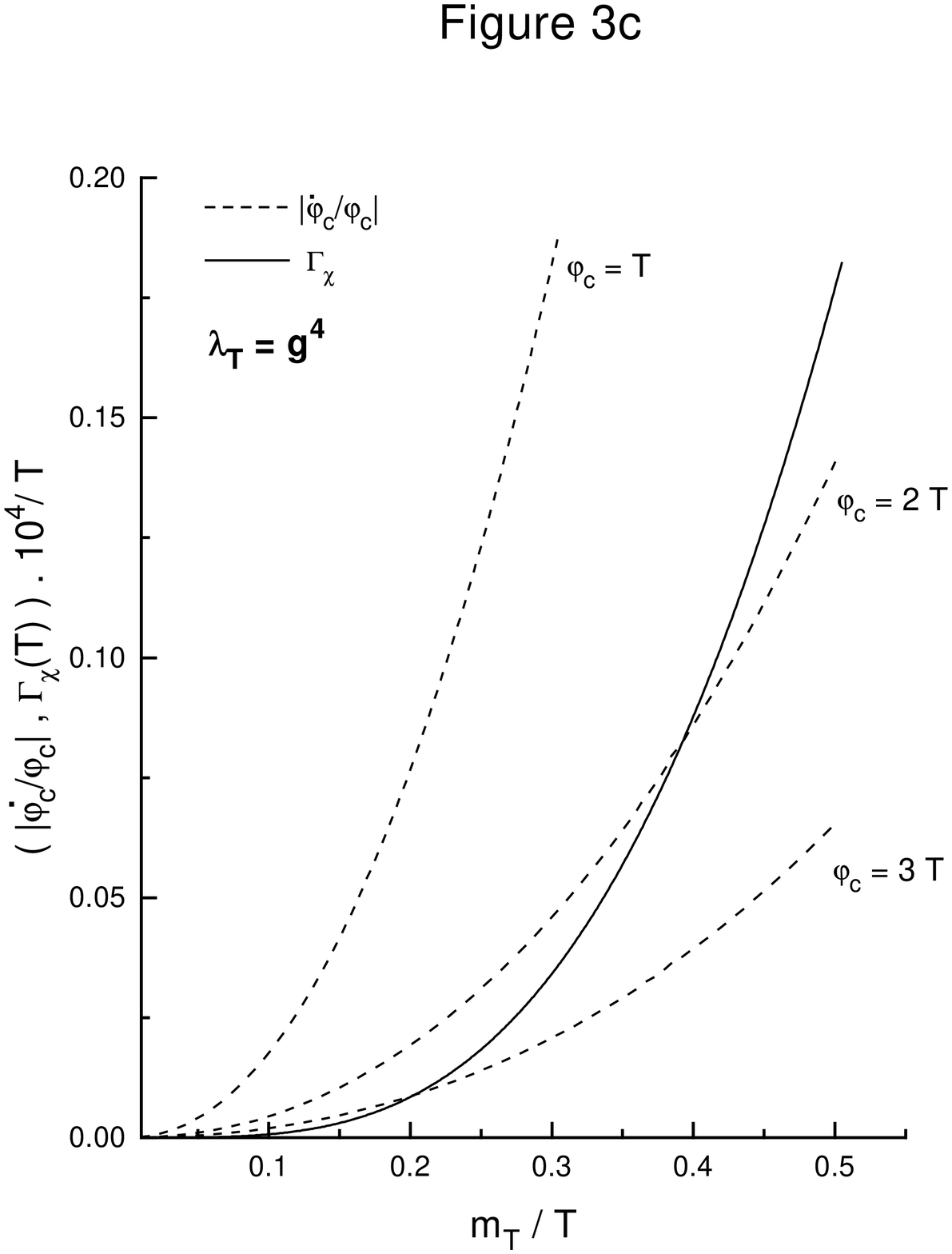}}}

\end{figure}

\end{document}